\documentclass[
twocolumn,
nofootinbib,
amsmath,amssymb,
floatfix,
pre,
floatfix,
superscriptaddress
]{revtex4-2}

\usepackage{graphicx,color}
\definecolor{brown}{rgb}{0.63,0.27,0.18}
\definecolor{orange}{rgb}{1.00,0.65,0.00}

\usepackage{afterpage} 

\usepackage{moresize}
\usepackage{dcolumn}
\usepackage{bm,multirow}

\makeatletter
\newcommand*{\balancecolsandclearpage}{%
  \close@column@grid
  \twocolumngrid
}
\makeatother

\newcommand{\be}{\begin{equation}}
\newcommand{\ee}{\end{equation}}
\usepackage{color}

\begin{document}

\newcommand {\rsq}[1]{\left< R^2 (#1)\right.}
\newcommand {\rsqL}{\left< R^2 (L) \right>}
\newcommand {\rsqbp}{\left< R^2 (N_{bp}) \right>}
\newcommand {\Nbp}{N_{bp}}
\newcommand {\etal}{{\em et al.}}
\newcommand{\Ham}{{\cal H}}
\newcommand{\AngeloComment}[1]{\textcolor{red}{(AR) #1}}
\newcommand{\MattiaComment}[1]{\textcolor{green}{(MU) #1}}

\newcommand{\NewText}[1]{\textcolor{orange}{#1}}
\newcommand{\scs}{\ssmall}

\newcommand{\Tau}{\mathrm{T}}



\title{Ring polymers in two-dimensional melts double-fold around randomly branching ``primitive shapes''} 

\author{Mattia Alberto Ubertini}
\email{mattia.ubertini@fmi.ch}
\affiliation{Friedrich Miescher Institute for Biomedical Research (FMI), 4056 Basel, Switzerland}

\author{Angelo Rosa}
\email{anrosa@sissa.it}
\affiliation{Scuola Internazionale Superiore di Studi Avanzati (SISSA), Via Bonomea 265, 34136 Trieste, Italy}

\date{\today}

\begin{abstract}
Drawing inspiration from the concept of the ``primitive path'' of a linear chain in melt conditions, we introduce here a numerical protocol which allows us to detect, in an unambiguous manner, the ``primitive shapes'' of ring polymers in two-dimensional melts.
Then, by analysing the conformational properties of these primitive shapes, we demonstrate that they conform to the statistics of two-dimensional branched polymers (or, trees) in the same melt conditions, in agreement with seminal theoretical work by Khokhlov, Nechaev and Rubinstein.
Results for polymer dynamics in light of the branched nature of the rings are also presented and discussed.
\end{abstract}

\maketitle

\section{Introduction}\label{sec:Intro}
\begin{figure}
\includegraphics[width=0.42\textwidth]{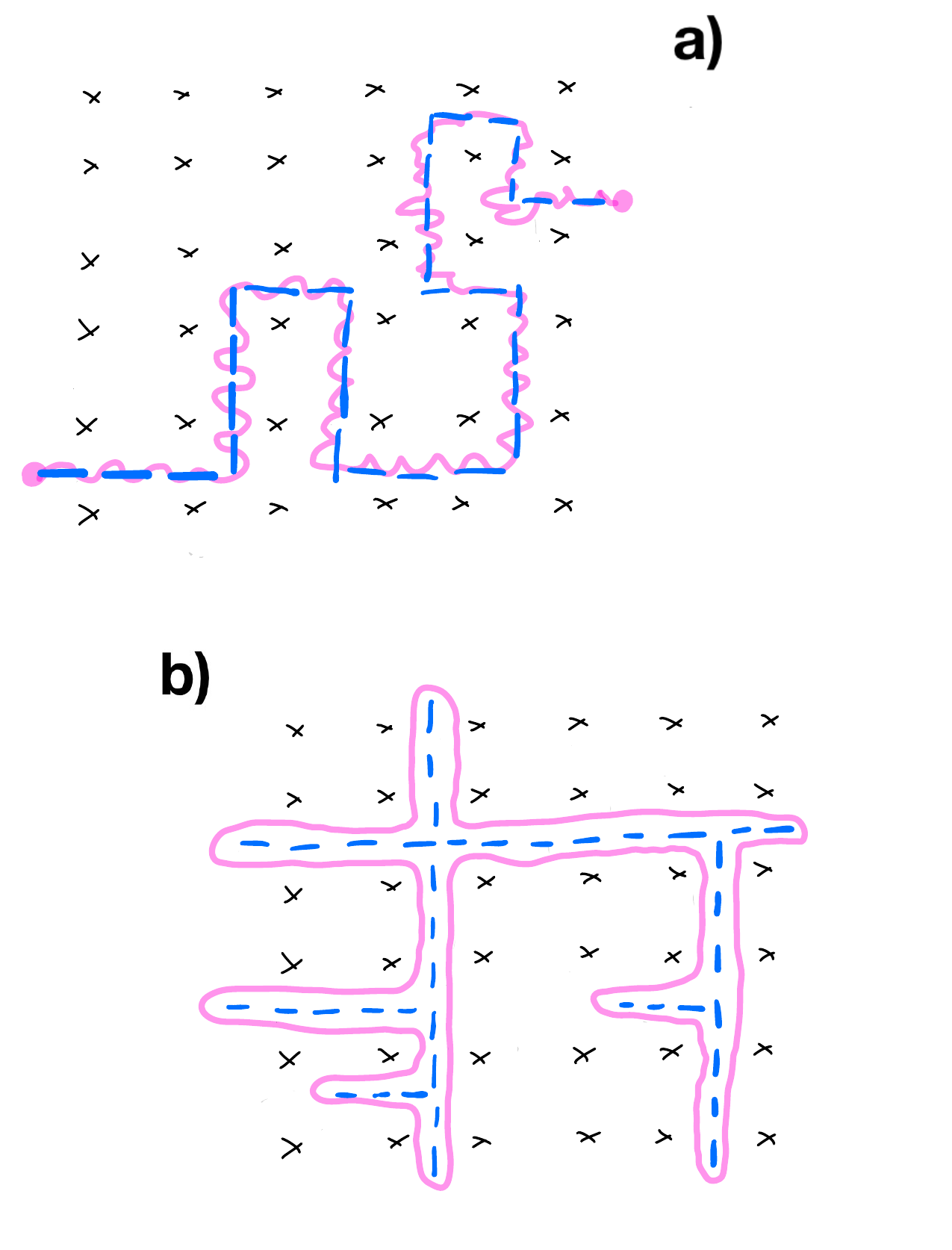}
\caption{
(a)
Classical illustration of the primitive path (dashed line) of a linear chain (solid line) in an entangled polymer melt.
By representing the effects of the constraints exerted by neighboring chains as a network of uncrossable topological obstacles ($\times$), the primitive path is identified as the shortest path which the polymer chain can be contracted into, at fixed chain ends and without crossing the obstacles.
(b)
In a melt of unknotted and non-concatenated rings, the equivalent topological barriers induce each ring to double-fold around a randomly branching path.
}
\label{fig:PPlinearsVSrings}
\end{figure}

Edwards introduced the concept of a ``primitive path'' long ago~\cite{Edwards1967} to explain the viscoelastic behavior of {\it entangled} polymer solutions and melts (see panel (a) in Figure~\ref{fig:PPlinearsVSrings}).
This idea has since become a cornerstone of polymer physics.
In a polymer melt the motion of each chain is constrained by its neighbors, which perform as uncrossable, {\it topological} obstacles caging the chain within a tube-like region~\cite{deGennes1971}.
The primitive path represents the central axis of this tube, and it is identified as the shortest path which the polymer chain can be contracted into, at fixed chain ends and without crossing the topological obstacles represented by the surrounding chains or each other~\cite{EveraersScience2004,ReadLikhtman2008}.
This definition, when applied to computer-generated conformations of entangled linear polymer melts~\cite{UchidaJCP2008}, allows researchers to map a melt of polymer chains into a melt of primitive paths.
Quite remarkably, this mapping provides a universal framework~\cite{EveraersCommodity2020} for describing the macroscopic rheological properties of entangled polymer liquids, regardless of their specific chemical composition.

The conventional definition of a primitive path is intuitive and simple, but it relies on the presence of free chain ends.
This raises a problem: can this concept be generalized to entangled polymers with no free ends?

In this work we explore this question by considering melts of unknotted and non-concatenated ring polymers, a particular class of systems that has challenged theorists~\cite{NechaevKhokhlov1985,Rubinstein1986,NechaevKoleva1987,Rubinstein1994,BreretonVilgis1995,KholodenkoVilgis1998,SakauePRL2011,ObukhovWittmer2014,GrosbergSM2014,PanyukovRubinsteinMacromolecules2016,Ferrari2019}, experimentalists~\cite{Kapnistos2008,Pasquino2013,Kruteva2017,KrutevaACSML2020,KrutevaPRL2020,Kruteva2023} and computational~\cite{CatesDeutschJPhysFrance1986,Cates-PRE1996,Cates-PRE2000,halverson2011molecular-statics,halverson2011molecular-dynamics,RosaEveraersPRL2014,SchramRosaEveraers2019,UbertiniSmrekRosa2022} physicists for decades.
In particular, Khokhlov and Nechaev~\cite{NechaevKhokhlov1985} and Rubinstein and coworkers~\cite{Rubinstein1986,Rubinstein1994} suggested that, because of the constraint of non-concatenation, rings minimize their mutual overlap by {\it double-folding} around {\it randomly branching}, or tree-like, primitive paths (see panel (b) in Fig.~\ref{fig:PPlinearsVSrings}).

The main goal of this work is to provide a direct numerical test of the Khokhlov-Nechaev-Rubinstein picture.
Specifically, we introduce a numerical algorithm to straightforwardly and unambiguously extract the primitive path (to which, to distinguish it from the primitive path of linear chains, we prefer from now on the term ``primitive shape'') of rings in two-dimensional computer-generated melts.
Our results show that these primitive shapes exhibit complex, non-trivial behavior which is, indeed, strongly reminiscent of branched, tree-like conformations.
To demonstrate that, we present a detailed characterization of their statistical properties using the set of observables and distribution functions previously introduced by one of us (A.R.) and R. Everaers~\cite{RosaEveraersJPA2016,RosaEveraersJCP2016,RosaEveraers-TreesPDF2017} for studying two- and three-dimensional lattice trees in various dilution conditions.

The paper is organized as follows.
In Section~\ref{sec:PolymerModel} we introduce the polymer model and the kinetic Monte Carlo algorithm used for the numerical simulations of ring melts, in Sec.~\ref{sec:SimulationDetails} we provide details on the simulated systems, while in Sec.~\ref{sec:RingsPrimitiveShape} we describe the algorithm used to reconstruct the primitive shape of each ring.
Sec.~\ref{sec:Theory-2dimBPs} is devoted to recapitulate the relevant state-of-the-art theoretical ideas describing the structure of interacting random trees, focusing in particular on observables (Sec.~\ref{sec:Theory-2dimBPs-Obs}) and distribution functions (Sec.~\ref{sec:Theory-2dimBPs-PDFs}).
In Sec.~\ref{sec:Results} we present the main results, where we frame the extracted primitive shapes of rings in terms of the theoretical ideas for $2d$ melts of random trees.
Then (Sec.~\ref{sec:Results-Dynamics}), we discuss these results at the light of ring dynamics, highlighting in particular the fact that $2d$ rings represent a quite special situation with respect to $3d$ rings.
Outlook and conclusions are presented in Sec.~\ref{sec:Concls}.

\section{Polymer model, simulation methods, analysis of polymer conformations}\label{sec:ModelMethods}

\subsection{Kinetic Monte Carlo algorithm}\label{sec:PolymerModel}

%
\begin{figure}
\includegraphics[width=0.42\textwidth]{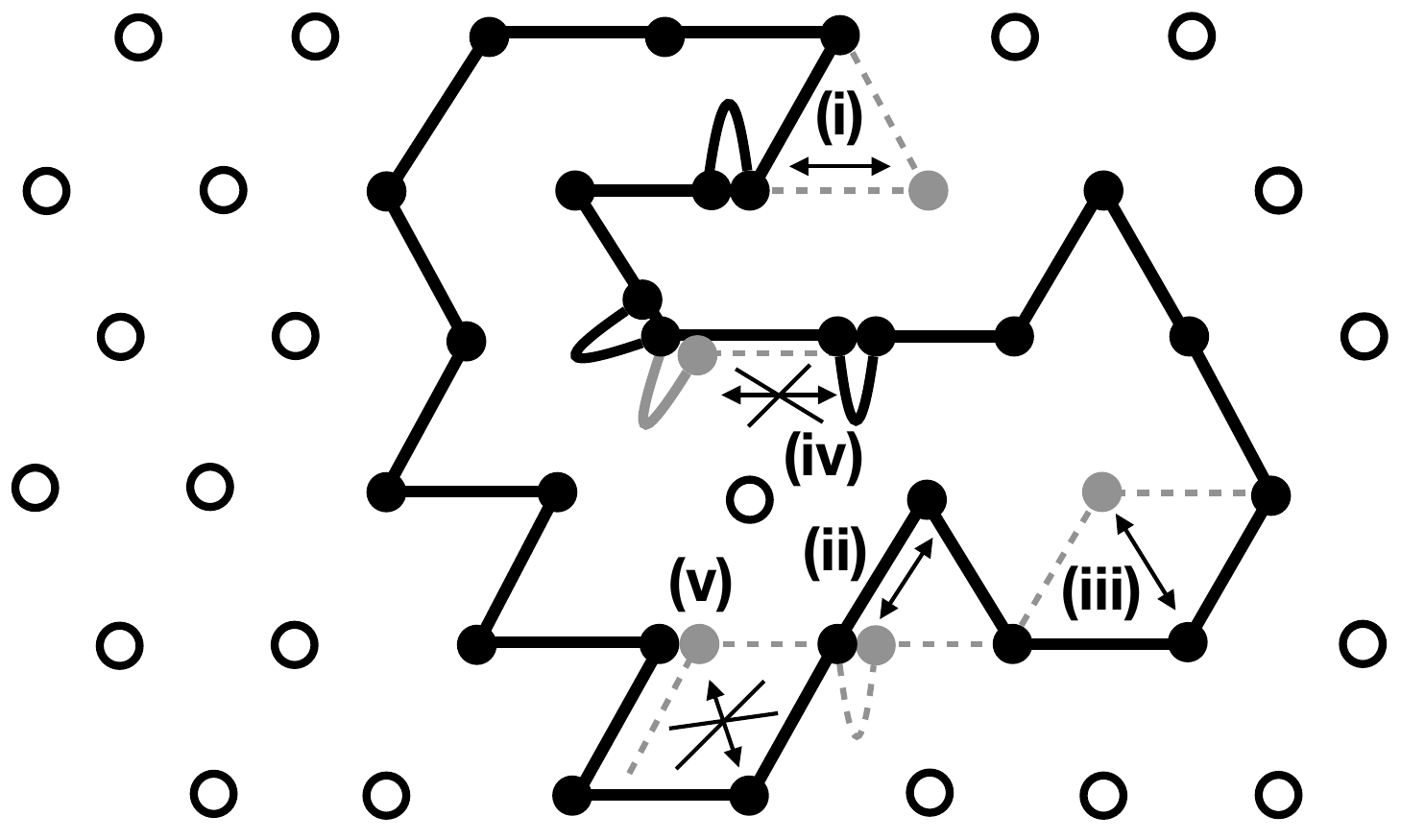}
\caption{
Two-dimensional illustration of the lattice polymer model and kinetic Monte Carlo moves.
Monomers (filled dots) occupy the spatial positions of the regular triangular lattice (empty dots) and two nearest neighbor monomers are connected by a black line representing the polymer bond between them.
For two nearest neighbor monomers occupying the same lattice site the bond that connects them (represented by an arc) represents a unit of ``stored length''.
Lattice positions connected by the double arrows are examples of {\it allowed} MC moves:
(i) a unit of stored length unfolding to a normal bond,
(ii) a bond folding into a unit of stored length,
(iii) a Rouse-like move.
Lattice positions connected by the double arrows with the cross are examples of {\it forbidden} MC moves:
(iv) a monomer moving to the left that would occupy a lattice position with two monomers already present,
(v) two non-nearest neighbor monomers on the same lattice site violating the excluded volume constraint.
}
\label{fig:MC_move}
\end{figure}

We adapt the efficient kinetic Monte Carlo (MC) algorithm employed in~\cite{Hugouvieux2009,Schram-LatticeModel2018,ubertini2021computer,UbertiniSmrekRosa2022} to construct equilibrated dense melts of self-avoiding ring polymers on the two-dimensional triangular lattice (see Figure~\ref{fig:MC_move}).
The lattice spacing, denoted by $a$, is chosen as the reference unit of length.
The algorithm, based on the elastic lattice polymer model analogous to the Rubinstein's repton model~\cite{Rubinstein-Repton-PRL1987}, works as the following.
Two consecutive monomers along the chain can either sit on nearest neighbour lattice sites or occupy the same lattice site, with strictly no more than two monomers on the same site.
At the same time, double occupancy of the same site by non-consecutive monomers is not permitted due to excluded volume.
As such, the bond length $b$ between nearest neighbor monomers takes the two possible values, $=a$ or $=0$, and in the latter case the bond is said to host a unit of {\it stored length}.
For a polymer with $N$ bonds, the total contour length $L \equiv N\langle b\rangle < Na$ where $\langle b\rangle$ is the average bond length.
This numerical ``trick'' makes the polymer elastic.

To explore the effects of chain composition, we have considered polymer chains of different stiffnesses which have been modelled by introducing the following energy penalty term~\cite{UbertiniSmrekRosa2022},
\begin{equation}\label{eq:CosPotential}
\frac{\mathcal H_{\rm bend}}{k_BT} = -\kappa_{\rm bend} \sum_{i=1}^{L / a} \cos \theta_i \equiv -\kappa_{\rm bend} \sum_{i=1}^{L / a} \frac{\vec{t}_i \cdot \vec{t}_{i+1}}{|\vec{t}_i| |\vec{t}_{i+1}|} \, ,
\end{equation}
for bending, where
$\kappa_{\rm bend}$ represents the bending stiffness parameter 
and
$\vec t_i \equiv \vec r_{i+1}-\vec r_{i}$ is the oriented bond vector between monomers $i$ and $i+1$ having spatial coordinates $\vec r_i$ and $\vec r_{i+1}$.\footnote{For ring polymers, it is implicitly assumed the periodic boundary condition along the chain $N+1\equiv 1$.}
It should be noticed that, since bond vectors are obviously ill-defined when two monomers form a stored length, the sum in Eq.~\eqref{eq:CosPotential} is restricted to the {\it effective} bonds of the chains.
By increasing $\kappa_{\rm bend}$, the energy term Eq.~\eqref{eq:CosPotential} makes polymers stiffer. 

Ring polymers are let to evolve in time, and thus reach equilibrium, by proposing moves that obey the occupancy constraints described above.
In more detail, a single move consists in randomly picking a monomer of one of the chains in the system and attempting its displacement towards one of the nearest lattice sites (see Fig.~\ref{fig:MC_move} for an illustration of possible situations that may occur). 
The move is accepted if chain connectivity is preserved and with the additional constraints that the destination lattice site is either (1) empty or (2) occupied at most by one of the nearest neighbor monomers along the chain.
In analogy with classical~\cite{DoiEdwardsBook,RubinsteinColbyBook,ubertini2021computer} polymer dynamics, case (1) is an example of {\it Rouse}-like move while case (2) is a {\it reptation}-like move (essentially the move produces mass drift along the contour length of the chain, as occurring in reptation dynamics).
By these two moves only, the algorithm reproduces known~\cite{DoiEdwardsBook,RubinsteinColbyBook} features of polymer dynamics and, thanks to the stored length ``trick'' which integrates local fluctuations of the chain density, remains efficient even when it is applied to the equilibration of very dense systems~\cite{Schram-LatticeModel2018}.

\subsection{Simulation details}\label{sec:SimulationDetails}
We have considered bulk solutions of $M$ closed (ring) polymer chains, each chain made of $N_{\rm ring}$ monomers or bonds, with values
$N_{\rm ring} \times M = \left[ 40 \times 1920, 80 \times 960, 160 \times 480, 320 \times 240, 640 \times 120, 1280 \times 60\right]$, with constant total number of monomers $=76'800$.
Bulk conditions are implemented through the enforcement of periodic boundary conditions in a simulation box of total surface $S=L^2$, where the linear sizes of the box, $L$, has been fixed based on the monomer number density $\rho a^2 \equiv \frac{N_{\rm ring} M}Sa^2 = 1.25$ that guarantees melt conditions.
All these systems have been studied for bending stiffness parameters $\kappa_{\rm bend} = 0, 1, 1.5$ (Eq.~\eqref{eq:CosPotential}).

Each system is initialized by arranging the rings inside the simulation box in a conformation that avoids unphysical overlaps between monomers.
Then, for each MC step we pick a single monomer at random, displace it towards a random lattice neighbor and check if the move satisfies the constraints as described in Sec.~\ref{sec:PolymerModel}.
By taking as unit of ``time'' $1 \, \mbox{MC sweep} \equiv (N_{\rm ring} \times M) \, \mbox{MC steps} = 76'800 \, \mbox{MC steps}$, we run simulations for $\simeq 10^7$ up to $\simeq 10^8$ MC sweeps.
Then, we monitor chain relaxation to equilibrium in terms of the monomer mean-square displacement relative to the chain center of mass~\cite{KremerGrest-JCP1990},
\begin{equation}\label{eq:Introduce-g2}
g_2(t) \equiv \left\langle \frac1{N_{\rm ring}} \sum_{i=1}^{N_{\rm ring}} ( \vec r_i(t) - \vec r_{\rm cm}(t) - \vec r_i(0) + \vec r_{\rm cm}(0) )^2 \right\rangle \, ,
\end{equation}
where ${\vec r}_{\rm cm} = \frac1{N_{\rm ring}} \sum_{i=1}^{N_{\rm ring}} {\vec r}_i$.
Asymptotically, $g_2(t\to\infty) = 2\langle R_g^2\rangle_{\rm ring}$ where
\begin{equation}\label{eq:<Rg2>_ring}
\langle R_g^2\rangle_{\rm ring} = \left\langle \frac1{N_{\rm ring}}\sum_{i=1}^{N_{\rm ring}} (\vec r_i-\vec r_{\rm cm})^2 \right\rangle
\end{equation}
is the ring mean-square gyration radius.
The appearance of a plateau in the large-time behavior of $g_2$ (see Fig.~\ref{fig:g2Plot} in Supplemental Material (SM)~\cite{SMnote}) is the signature that our chains have attained complete structural relaxation.

\subsection{Reconstructing the primitive shapes of ring polymers: algorithmic definition}\label{sec:RingsPrimitiveShape}
We reconstructed the primitive shapes of individual ring polymer configurations in a $2d$ melt configuration by introducing a numerical algorithm based on the original intuitions of Khokhlov and Nechaev~\cite{NechaevKhokhlov1985} and Rubinstein~\cite{Rubinstein1986,Rubinstein1994}, according to which each ring polymer behaves like being confined in a lattice of fixed obstacles (representing all the surrounding rings in the melt) and it can be mapped then onto a tree-like structure on the corresponding {\it dual} lattice where such obstacles are absent.

\begin{figure*}
\includegraphics[width=0.95\textwidth]{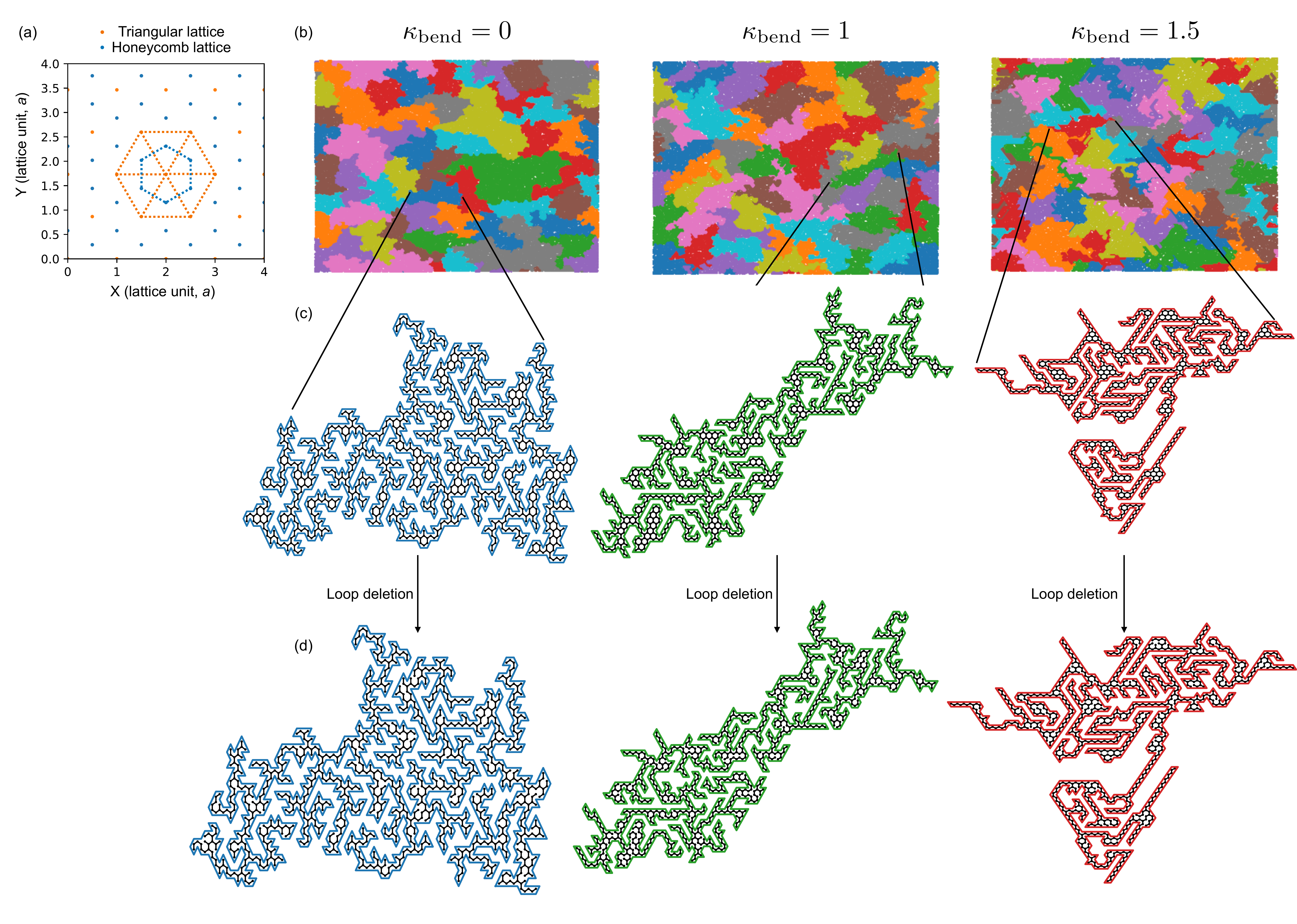}
\caption{
(a)
Illustration of the $2d$ triangular lattice (orange dots, with $6$ unit cells drawn explicitly) of unit step $=a$ where ring polymers are simulated, along with the dual honeycomb lattice (blue dots, with $1$ cell drawn explicitly) of unit step $=a/\sqrt{3}$ where the corresponding trees reside.
(b)
Snapshots of ring melt configurations for $N_{\rm ring} =1280$ and $\kappa_{\rm bend} = 0, 1, 1.5$ (from left to right, see legend).
(c)
Examples of single ring polymers (blue/green/red circular contours) isolated from their corresponding melts, along with their primitive tree-like backbones (black lines inside each circular contour).
Notice that the trees contain a certain amount of loops.
(d)
Final, loop-less, primitive shapes obtained by randomly removing one bond in each of the former loops.
}
\label{fig:Primitive-Path}
\end{figure*}

In our simulations polymers reside on the $2d$ triangular lattice of reference unit step $=a$, whose dual is the honeycomb lattice (panel (a) of Fig.~\ref{fig:Primitive-Path}) of unit step $=a/\sqrt{3}$.
For each ring in a typical melt configuration (panel (b) of Fig.~\ref{fig:Primitive-Path}), we identified first the honeycomb lattice sites residing {\it within} its contour which allow us to identify, without ambiguity, the nodes of the associated ``backbone''.
Then, we connect any two of these nodes of the honeycomb lattice by an edge whenever they are spatial nearest neighbors.
Visual inspection on the obtained conformations (black lines in panel (c) of Fig.~\ref{fig:Primitive-Path}) confirms indeed a pronounced tree-like architecture.

If a ring was perfectly double-folded, the resulting tree backbone would be, by construction, without closed loops.
In our simulated structures, a non-null amount of loops per ring is however always found (Fig.~\ref{fig:Primitive-Path}(c)).
On the other hand, after easily verifying that such an amount remains small, essentially one may assume that loops do not compromise in any fundamental manner the statistical description of the trees, only they would make the analysis of tree conformations technically harder.
For these reasons, we introduce the rule of removing loops simply by randomly cutting one bond of each of them (panel (d) of Fig.~\ref{fig:Primitive-Path}).
As anticipated the fraction of removed bonds is small, at most $\simeq 8\%$ for $\kappa_{\rm bend} = 1.5$ of the original looped conformations (see Fig.~\ref{fig:Statistics_Bond_Removed} in SM~\cite{SMnote}).

\subsection{Trees analysis}\label{sec:TreeAnalysis}
Each reconstructed tree is made of a total number of ``honeycomb-lattice'' nodes $N_{\rm tree}$ which, on average, is equal to $\simeq 80\%$ of the original number of ``triangular-lattice'' nodes $N_{\rm ring}$ of each ring (l.h.s panel of Fig.~\ref{fig:TreeMassVSRingMass} in SM~\cite{SMnote}).

As a consequence of the properties of the honeycomb lattice, every tree node may have one, two or, at most, three bonds protruding from it;\footnote{As noticed in works~\cite{RosaEveraersJPA2016,RosaEveraersJCP2016,RosaEveraers-TreesPDF2017,vdHoek2024}, there is no substantial loss of generality by choosing a model for branched polymers where nodes host no more than three bonds compared to models featuring nodes with more than three bonds protruding from them.} in the last case, we call that node a ``branch-node''.
Interestingly, we have found that the mean fraction of branch-nodes depends on the original ring stiffness $\kappa_{\rm bend}$, with flexible rings being sensibly {\it more} branched than stiffer rings.
Indeed, as shown in the r.h.s. panel of Fig.~\ref{fig:TreeMassVSRingMass} in SM~\cite{SMnote}, the amount of branch-nodes saturates in the asymptotic limit of large rings ({\it i.e.}, trees) from $\simeq 22\%$ for $\kappa_{\rm bend}=0$ up to $\simeq 13\%$ for $\kappa_{\rm bend}=1.5$.

After reconstructing the coordinates and bond connectivities of the ring primitive shapes, we have analyzed tree connectivity and spatial structure by using the ``burning'' algorithm for trees originally introduced in Ref.~\cite{RosaEveraersJPA2016}.
This algorithm works iteratively: each step consists in removing from the list of all tree nodes those with only one bond protruding from it and updating the number of bonds and the indices of the remaining ones accordingly.
The algorithm stops when only one node (the so called ``center'' of the tree) remains in the list.
In this way, by keeping track of the nodes that have been removed, it is possible to obtain information about the mass and the shape of tree branches (Sec.~\ref{sec:Theory-2dimBPs-Obs-Defs}).
The algorithm can be then generalized to detect also the minimal path length $\ell_{ij}$ between any pair of nodes $i$ and $j$ (Sec.~\ref{sec:Theory-2dimBPs-Obs-Defs}): it is in fact sufficient that both nodes ``survive'' the burning process.
For a more detailed illustration of the burning algorithm and of its applications on trees, see Refs.~\cite{RosaEveraersJPA2016,RosaEveraersJCP2016,RosaEveraers-TreesPDF2017}.

\section{Structure of two-dimensional melts of branched polymers: theory and methods}\label{sec:Theory-2dimBPs}
In this Section, we introduce quantities and methods of analysis that will be used (in Sec.~\ref{sec:Results}) to characterize the tree-like conformational properties of the ring primitive shapes.

\subsection{Observables}\label{sec:Theory-2dimBPs-Obs}

\subsubsection{Definitions}\label{sec:Theory-2dimBPs-Obs-Defs}
To explore the connection between the primitive shape of our $2d$ rings and the physics of branched polymers, we follow closely the works~\cite{RosaEveraersJPA2016,RosaEveraersJCP2016} and discuss the power-law behaviors for the following observables:
\begin{enumerate}
\item
The mean path length as a function of the mean tree weight $\langle N_{\rm tree}\rangle$:\footnote{In the mentioned works~\cite{RosaEveraersJPA2016,RosaEveraersJCP2016} $N_{\rm tree}$ is not a fluctuating quantity. Here, as shown in the l.h.s. panel of Fig.~\ref{fig:TreeMassVSRingMass} in SM~\cite{SMnote}, the relative fluctuations around the mean value $\langle N_{\rm tree}\rangle$ become rapidly small in the large-$N_{\rm ring}$ limit, so they are expected to have a negligible effect on the estimation of scaling exponents.}
\begin{equation}\label{eq:L-Nrho}
\langle L\rangle \sim \langle N_{\rm tree}\rangle^\rho \, ,
\end{equation}
where $L \equiv \frac1{N_{\rm tree}(N_{\rm tree}-1)} \sum_{i=1}^{N_{\rm tree}-1}\sum_{j=i+1}^{N_{\rm tree}} \ell_{ij}$ and $\ell_{ij}$ is the minimal path length connecting the pair of nodes $i$ and $j$ on a tree of weight $=N_{\rm tree}$.
\item
The mean branch weight as a function of the mean tree weight $\langle N_{\rm tree}\rangle$:
\begin{equation}\label{eq:Nbr-Neps}
\langle N_{\rm br} \rangle \sim \langle N_{\rm tree}\rangle^\epsilon \, ,
\end{equation}
where $N_{\rm br} \equiv \min(n, N_{\rm tree}-1-n)$ is the weight of the smallest of the two sub-trees that are obtained by cutting one tree bond.
\item
The mean-square end-to-end spatial distance as a function of the mean path length $\langle L\rangle$:
\begin{equation}\label{eq:R2ell}
\langle R_{\rm path}^2 \rangle \sim \langle L\rangle^{2\nu_{\rm path}}
\end{equation}
\item
The tree mean-square gyration radius as a function of the mean tree weight $\langle N_{\rm tree}\rangle$:
\begin{equation}\label{eq:Rg2-N2nu}
\langle R_g^2\rangle \sim \langle N_{\rm tree}\rangle^{2\nu} \, ,
\end{equation}
where $R_g^2 \equiv \frac1{N_{\rm tree}}\sum_{i=1}^{N_{\rm tree}} (\vec r_i-\vec r_{\rm cm})^2$ and $\vec r_{\rm cm} = \frac1{N_{\rm tree}} \sum_{i=1}^{N_{\rm tree}} \vec r_i$ is the spatial position of the tree center of mass.
\end{enumerate}
Notice the fundamental relations $\rho = \epsilon$~\cite{Rensburg1992} and $\nu = \rho \, \nu_{\rm path}$~\cite{RosaEveraersJPA2016,RosaEveraersJCP2016} holding between exponents.
The complete list of the measured mean values with corresponding error bars for the different ring systems is presented in Table~\ref{tab:TreeObsMeanValues} in SM~\cite{SMnote}.

\begin{table*}
\begin{tabular}{ccccccc}
& \, \, Relation to \, \, & \, \, Measured value \, \, & \multicolumn{4}{c}{\, \, Measured value (this work) \, \,} \\
& \, \, other exponents~\cite{RosaEveraersJCP2016} \, \, & (Ref.~\cite{RosaEveraersJCP2016}) & \, $\kappa_{\rm bend}=0$ \, & \, $\kappa_{\rm bend}=1$ \, & \, $\kappa_{\rm bend}=1.5$ \, & \\
\hline
\hline
\\
$\langle L\rangle \sim \langle N_{\rm tree}\rangle^{\rho}$ & -- & $0.613 \pm 0.007$ & $0.63 \pm 0.04$ & $0.61 \pm 0.05$ & $0.59 \pm 0.07$ & \\
\\
$\langle N_{\rm br}\rangle \sim \langle N_{\rm tree}\rangle^{\epsilon}$ & $\epsilon = \rho$ & $0.63 \pm 0.01$ & $0.64 \pm 0.05$ & $0.63 \pm 0.04$ & $0.63 \pm 0.03$ & \\
\\
$\langle R_{\rm path}^2\rangle \sim \langle L\rangle^{2\nu_{\rm path}}$ & -- & $0.780 \pm 0.005$ & $0.783 \pm 0.006$ & $0.79 \pm 0.02$ & $0.687 \pm 0.006$ & \\
\\
$\langle R_g^2\rangle \sim \langle N_{\rm tree}\rangle^{2\nu}$ & $\nu = \rho \, \nu_{\rm path}$ & $0.48 \pm 0.02$ & $0.50 \pm 0.02$ & $0.499 \pm 0.003$ & $0.461 \pm 0.002$ & \\
\\
\hline
\end{tabular}
\caption{
Exponents characterizing tree observables (column 1) and relations between them (column 2, see Ref.~\cite{RosaEveraersJCP2016} for details).
Column 3 reports the values for $2d$ tree melts estimated in Ref.~\cite{RosaEveraersJCP2016}.
Columns 4 to 6 report the estimated values for the tree-like primitive shapes of the $2d$ ring melts considered in this work, for different ring bending stiffness $\kappa_{\rm bend}$.
For technical details on the derivation of these exponents, see Sec.~\ref{sec:Theory-2dimBPs-Obs-Methods}.
}
\label{tab:ExpSummary-Obs}
\end{table*}

\subsubsection{Extrapolation of exponents from data for finite-size trees}\label{sec:Theory-2dimBPs-Obs-Methods}
In order to get reliable values and errors of scaling exponents ``$\rho, \epsilon, \nu_{\mathrm{path}}, \nu$'' in the asymptotic ({\it i.e.}, $\langle N_{\rm tree}\rangle \to \infty$) limit, we adopt the following strategy, similar to the one employed in previous works~\cite{RosaEveraersJPA2016,RosaEveraersJCP2016}, which combines together the results obtained from fitting the $\langle N_{\rm tree}\rangle$-dependent data to two functional forms.

To fix the ideas, take the data for the observable $\langle R_g^2\rangle$ as a function of $\langle N_{\rm tree}\rangle$ and the corresponding exponent $\nu$ and use the following expressions:
\begin{enumerate}
\item
A simple power-law behavior with two (c, $\nu$) fit parameters:
\begin{equation}\label{eq:Fit-2params}
\log( \langle R_g^2 \rangle ) = 2\nu \log( \langle N_{\rm tree} \rangle ) + c \, ,
\end{equation}
which we use only on the data with $N_{\rm ring} \geq320$ (namely for the three largest rings/trees).
\item
A power-law behavior with a correction-to-scaling term and four ($c$, $d$, $\Delta$, $\nu$) fit parameters:
\begin{equation}\label{eq:Fit-4params}
\log( \langle R_g^2 \rangle ) = 2 \nu \log(\langle N_{\rm tree} \rangle) + c + \frac{d}{\langle N_{\rm tree} \rangle^{\Delta}} \, ,
\end{equation}
which we use on the entire $N_{\rm ring}$-range.
\end{enumerate}
For the other exponents, analogous expressions to Eq.~\eqref{eq:Fit-2params} and Eq.~\eqref{eq:Fit-4params} have been employed.
In all cases, best fit parameters were obtained by standard $\chi^2$-minimization. 
The results of the two- (Eq.~\eqref{eq:Fit-2params}) and four-parameter (Eq.~\eqref{eq:Fit-4params}) fits are collected in Table~\ref{tab:TreeObsFitResults} in SM~\cite{SMnote}.
Overall the two procedures lead to similar results for the scaling exponents; 
we combine then the two results together as: 
\begin{equation}\label{eq:FinalScalinExponents}
\nu = \frac{\nu({\rm Eq.}~\eqref{eq:Fit-2params}) + \nu({\rm Eq.}~\eqref{eq:Fit-4params})}2 \pm \frac{|\nu({\rm Eq.}~\eqref{eq:Fit-2params}) - \nu({\rm Eq.}~\eqref{eq:Fit-4params})|}2 \, ,
\end{equation}
to obtain a final estimate and the corresponding error bar.
See Table~\ref{tab:ExpSummary-Obs} for summary, and for comparison with the reference values of the exponents for $2d$ tree melts reported in Ref.~\cite{RosaEveraersJCP2016}.

\subsection{Distribution functions}\label{sec:Theory-2dimBPs-PDFs}

\subsubsection{Definitions}\label{sec:Theory-2dimBPs-PDFs-Defs}
As originally pointed out in~\cite{RosaEveraers-TreesPDF2017}, additional insight into the statistics of lattice trees can be gained by moving beyond the mere averages~\eqref{eq:Nbr-Neps}-\eqref{eq:Rg2-N2nu} and considering the related distribution functions that obey insightful scaling relations.
Specifically, in this work we will focus on:
\begin{enumerate}
\item
The distribution $p_{\langle N_{\rm tree}\rangle}(\ell)$ of path lengths $\ell$.
Here, data from different $\langle N_{\rm tree}\rangle$ are expected to superimpose, when expressed as functions of the scaled distances, $x = \ell / \langle L\rangle$:
\begin{equation}\label{eq:pNell}
p_{\langle N_{\rm tree}\rangle}(\ell) = \frac1{\langle L\rangle} \, q\!\left( \frac{\ell}{\langle L\rangle} \right) \, .
\end{equation}
Moreover, the function $q=q(x)$ is expected to be of the so called Redner-des Cloizeaux (RdC) form~\cite{Redner1980,DesCloizeauxBook},
\begin{equation}\label{eq:q_RdC_path}
q(x) =  C_{\ell} \, x^{\theta_{\ell}} \, \exp \left( -(K_{\ell} \, x)^{t_{\ell}} \right) \, ,
\end{equation}
where the exponents $(\theta_{\ell}, t_{\ell})$ depend on the trees universality class, while the numerical factors $(C_{\ell}, K_{\ell})$ are given by the analytical expressions,
\begin{eqnarray}
C_{\ell} & = & t_{\ell} \, \frac{\Gamma^{\theta_{\ell} + 1}((\theta_{\ell} + 2) / t_{\ell})}{\Gamma^{\theta_{\ell}+2}((\theta_{\ell}+1) / t_{\ell})} \, , \label{eq:RdC_C_l} \\
K_{\ell} & = & \frac{\Gamma((\theta_{\ell}+2) / t_{\ell})}{\Gamma((\theta_{\ell} + 1) / t_{\ell})} \, , \label{eq:RdC_K_l}
\end{eqnarray}
which can be found by imposing
(i) that $q(x)$ is normalized to $1$ ({\it i.e.}, $\int_0^\infty dx \, q(x)=1$)
and
(ii) that the {\it first moment} is the only scaling length ({\it i.e.}, $\langle x\rangle \equiv \int_0^\infty dx \, x q(x)=1$).
\item
The distribution, $p_{\langle N_{\rm tree}\rangle}(n)$, of branch sizes $n$.
For $1 \ll n \ll \langle N_{\rm tree}\rangle$, data from different trees are expected to superimpose and to display universal power-law behavior
\begin{equation}\label{eq:pNn}
p_{\langle N_{\rm tree}\rangle}(n) \sim n^{-(2-\epsilon)} \, .
\end{equation}
\item
The distribution $p_{\langle N_{\rm tree}\rangle}(\vec r | \langle L\rangle)$ of end-to-end spatial distances $\vec r$ of paths of length $=\langle L\rangle$ on trees of mass $\langle N_{\rm tree}\rangle$. 
The data superimpose, when expressed as functions of the scaled distances, $x = \left| \vec r \right| / \sqrt{\langle R_{\rm path}^2\rangle}$:
\begin{equation}\label{eq:pr_of_l}
p_{\langle N_{\rm tree}\rangle}(\vec r | \langle L\rangle) = \frac1{\langle R_{\rm path}^2\rangle} \, q\!\left( \frac{\left| \vec r \right|}{\sqrt{\langle R_{\rm path}^2\rangle}} \right) \, ,
\end{equation}
and the function $q=q(x)$ is expected to be of the following RdC form:
\begin{equation}\label{eq:q_RdC_e2e_ell}
q(x) = C_{\rm path} \, x^{\theta_{\rm path}} \, \exp \left( -(K_{\rm path} x)^{t_{\rm path}} \right) \, .
\end{equation}
\item
The distribution $p_{\langle N_{\rm tree}\rangle}(\vec r)$ of vector distances $\vec r$ between tree nodes.
Once again, data from different $\langle N_{\rm tree}\rangle$ are expected to superimpose when expressed as functions of the scaled distances, $x = \left| \vec r \right| / \sqrt{2 \langle R_g^2\rangle}$:
\begin{equation}\label{eq:pr}
p_{N_{\rm tree}}(\vec r) = \frac1{2 \langle R_g^2\rangle} \, q\!\left( \frac{\left| \vec r \right|}{\sqrt{2 \langle R_g^2\rangle}} \right) \, ,
\end{equation}
and, again, the function $q=q(x)$ is expected to be of the RdC form,
\begin{equation}\label{eq:q_RdC_tree}
q(x) =  C_{\rm tree} \, x^{\theta_{\rm tree}} \, \exp \left( -(K_{\rm tree} \, x)^{t_{\rm tree}} \right) \, .
\end{equation}
The two sets of exponents $(\theta_{\rm path}, t_{\rm path})$ (Eq.~\eqref{eq:q_RdC_e2e_ell}) and $(\theta_{\rm tree}, t_{\rm tree})$ (Eq.~\eqref{eq:q_RdC_tree}) depend on the tree universality class, while the two sets of numerical factors $(C_{\rm path}, K_{\rm path})$ and $(C_{\rm tree}, K_{\rm tree})$ are given by the analytical expressions,
\begin{eqnarray}
C & = & t \, \frac{\Gamma(1+\frac d2) \, \Gamma^{\frac{d+\theta}2}(\frac{2+d+\theta}t)}{d \, \pi^{d/2} \, \Gamma^{\frac{2+d+\theta}2}(\frac{d+\theta}t)} \, , \label{eq:RdC_C}\\
K & = & \sqrt{ \frac{\Gamma(\frac{2+d+\theta}t)}{\Gamma(\frac{d+\theta}t)} } \, , \label{eq:RdC_K}
\end{eqnarray}
obtained by imposing
(i) that $q(x)$ is normalized to $1$ ({\it i.e.}, $\int_0^\infty dx \, x q(x)=1$)
and
(ii) that the {\it second moment} is the only scaling length ({\it i.e.}, $\langle x^2 \rangle \equiv \int_0^\infty dx \, x^3 q(x)=1$).
\end{enumerate}
Notice that general scaling arguments~\cite{RosaEveraers-TreesPDF2017} imply the following fundamental relations between the exponents of RdC functions and the exponents characterizing observables:
$\theta_{\ell} = 1/\rho-1$,
$t_{\ell} = 1/(\rho-1)$,
$t_{\rm path} = 1/(1-\nu_{\rm path})$,
$\theta_{\rm tree} = \min(\theta_{\rm path}, 0)$,
$t_{\rm tree} = 1/(1-\nu)$.
The only independent exponent is $\theta_{\rm path}$, which turns out to be $>0$ based on numerical estimates.
Accordingly, $\theta_{\rm tree}=0$.
For a detailed characterization of lattice trees in terms of distribution functions and how these are related to RdC functions and their exponents, the reader is referred to Ref.~\cite{RosaEveraers-TreesPDF2017}.

\begin{table*}
\begin{tabular}{cccccc}
& \, \, Relation to \, \, & \, \, Measured value \, \, & \multicolumn{3}{c}{\, \, Measured value (this work) \, \,} \\
& \, \, other exponents~\cite{RosaEveraers-TreesPDF2017} \, \, & \, \, (Ref.~\cite{RosaEveraers-TreesPDF2017}) \, \, & \, $\kappa_{\rm bend}=0$ \, & \, $\kappa_{\rm bend}=1$ \, & \, $\kappa_{\rm bend}=1.5$ \, \\
\hline
\hline
\\
$\theta_{\ell}$ & $\frac1{\rho} - 1$ & $0.593 \pm 0.003$ & $0.421 \pm 0.001$ & $0.430 \pm 0.002$ & $0.413 \pm 0.003$ \\
$t_{\ell}$ & $\frac1{\rho-1}$ & $2.35 \pm 0.01$ & $2.631 \pm 0.006$ & $2.605 \pm 0.009$ & $2.57 \pm 0.01$ \\
\\
$\theta_{\rm path}$ & -- & $0.63 \pm 0.04$ & $0.679 \pm 0.001$ & $0.711 \pm 0.002$ & $0.6602 \pm 0.0004$ \\
$t_{\rm path}$ & $\frac1{1-\nu_{\rm path}}$ & $4.2 \pm 0.1$ & $6.132 \pm 0.006$ & $4.665 \pm 0.003$ & $2.94 \pm 0.02$ \\
\\
$\theta_{\rm tree}$ & $\min( \theta_{\rm path}, 0)=0$ & $-0.14 \pm 0.02$ & $-0.09 \pm 0.009$ & $-0.07 \pm 0.01$ & $-0.06 \pm 0.01$ \\
$t_{\rm tree}$ & $\frac1{1-\nu}$ & $1.857 \pm 0.005$ & $1.75 \pm 0.02$ & $1.72 \pm 0.03$ & $1.65 \pm 0.02$ \\
\\
\hline
\end{tabular}
\caption{
Exponents characterizing the Redner-des Cloizeaux behavior of distribution functions (column 1) and relations to other exponents (column 2, see Ref.~\cite{RosaEveraers-TreesPDF2017} for details).
Column 3 reports the values for $2d$ tree melts estimated in Ref.~\cite{RosaEveraers-TreesPDF2017}.
Columns 4 to 6 report the estimated values for the tree-like primitive shapes of the $2d$ ring melts considered in this work, for different ring bending stiffness $\kappa_{\rm bend}$.
For technical details on the derivation of these exponents, see Sec.~\ref{sec:Theory-2dimBPs-PDFs-Methods}.
}
\label{tab:ExpSummary-PDFs}
\end{table*}

\subsubsection{Extrapolation of exponents from data for finite-size trees}\label{sec:Theory-2dimBPs-PDFs-Methods}
For each $\langle N_{\rm tree}\rangle$, estimates of exponents and corresponding errors $(\theta_{\ell} \pm \Delta\theta_{\ell}, t_{\ell} \pm \Delta t_{\ell})$, $(\theta_{\rm path} \pm \Delta\theta_{\rm path}, t_{\rm path} \pm \Delta t_{\rm path})$ and $(\theta_{\rm tree} \pm \Delta\theta_{\rm tree}, t_{\rm tree} \pm \Delta t_{\rm tree})$ were obtained by best fit of RdC functions Eq.~\eqref{eq:q_RdC_path} (with Eqs.~\eqref{eq:RdC_C_l} and~\eqref{eq:RdC_K_l}), Eq.~\eqref{eq:q_RdC_e2e_ell} and Eq.~\eqref{eq:q_RdC_tree} (with Eqs.~\eqref{eq:RdC_C}-\eqref{eq:RdC_K}) to, respectively, data sets for $p_{\langle N_{\rm tree}\rangle}(\ell)$, $p_{\langle N_{\rm tree}\rangle}(\vec r | \langle L\rangle)$ and $p_{\langle N_{\rm tree}\rangle}(\vec r)$.
Results are collected in Table~\ref{tab:TreePDFsFitResults} in SM~\cite{SMnote}.

Then, to extrapolate the values and errors in the asymptotic regime $\langle N_{\rm tree}\rangle \to \infty$, we have fitted the straight line
\begin{equation}\label{eq:ExtractTheta-and-t}
y = -Ax + B
\end{equation}
to the data $(1/\langle N_{\rm tree}\rangle, \theta_{\ell})$, and so on for the other exponents, for the three largest $\langle N_{\rm tree}\rangle$'s and with fit parameters $A$ and $B$ (see Fig.~\ref{fig:RdC_estimatedparams_and_errors} in SM~\cite{SMnote} for a graphical illustration of the procedure).
The intercept with the $y$-axis, $B$, gives the extrapolated value, while error estimate of the extrapolated value is obtained by repeating this procedure to data sets $(1/\langle N_{\rm tree}\rangle, \theta_{\ell}+\Delta t_{\ell})$ and $(1/\langle N_{\rm tree}\rangle, \theta_{\ell}-\Delta t_{\ell})$ and so on for the other exponents.
Final results are summarized in Table~\ref{tab:ExpSummary-PDFs}.
For comparison, we report also the measured values of the exponents for $2d$ tree melts published originally in Ref.~\cite{RosaEveraers-TreesPDF2017}.

\section{Results}\label{sec:Results}

%
\begin{figure*}
$$
\begin{array}{ccc}
\mbox{(a)} & & \mbox{(b)} \\
\includegraphics[width=0.40\textwidth]{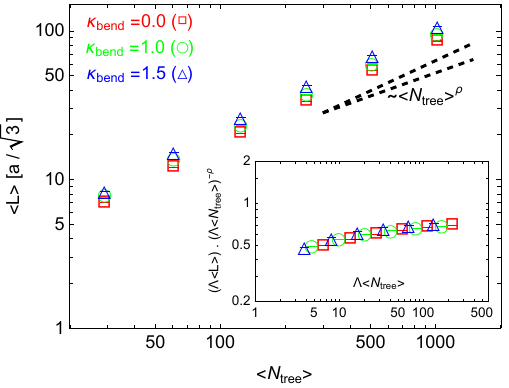} & & \includegraphics[width=0.40\textwidth]{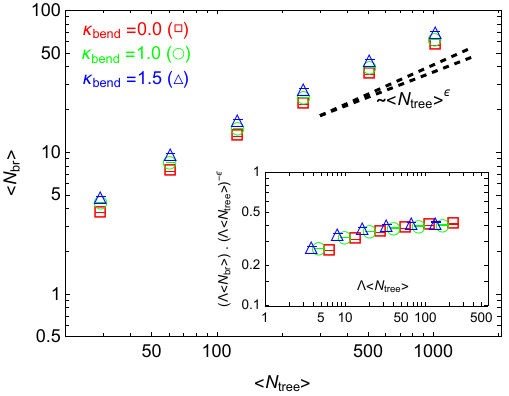} \\
\\
\mbox{(c)} & & \mbox{(d)} \\
\includegraphics[width=0.40\textwidth]{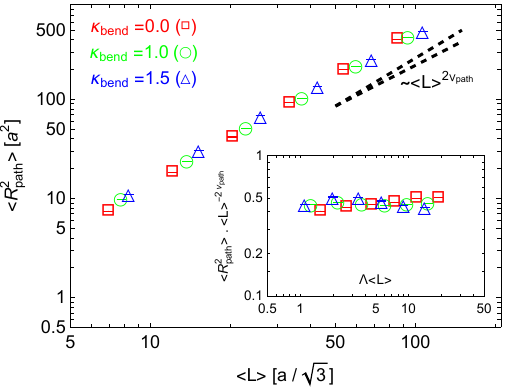} & & \includegraphics[width=0.40\textwidth]{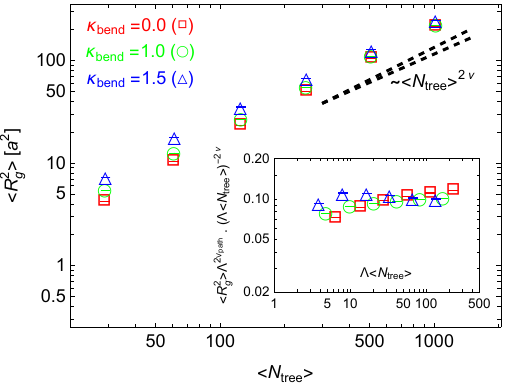} \\
\end{array}
$$
\caption{
Conformational properties of trees: observables (symbols) and asymptotic power-law behaviors (dashed lines).
(a)
$\langle L\rangle \sim \langle N_{\rm tree}\rangle^{\rho}$, mean path length as a function of the mean tree weight $\langle N_{\rm tree}\rangle$.
(b)
$\langle N_{\rm br}\rangle \sim \langle N_{\rm tree}\rangle^{\epsilon}$, mean branch weight as a function of the mean tree weight $\langle N_{\rm tree}\rangle$.
(c)
$\langle R_{\rm path}^2\rangle \sim \langle L\rangle^{2\nu_{\rm path}}$, mean-square end-to-end spatial distance of paths of length $=\langle L\rangle$.
(d)
$\langle R_g^2\rangle \sim \langle N_{\rm tree}\rangle^{2\nu}$, mean-square gyration radius as a function of the mean tree weight $\langle N_{\rm tree}\rangle$.
In each panel the dashed lines express the interval of possible values for the exponent, lying between the minimum lower-bound and the maximum upper-bound for all $\kappa_{\rm bend}$ (see Table~\ref{tab:ExpSummary-Obs}).
(Insets)
Universal scaling plots for the observables.
Here, the value of the exponent used in each plot corresponds to the average of the estimated best values for individual $\kappa_{\rm bend}$ (see Table~\ref{tab:ExpSummary-Obs}).
}
\label{fig:SummarizingObs}
\end{figure*}
%

\subsection{Observables}\label{sec:Results-Obs}
We start first with the mean path length $\langle L\rangle$ as a function of the mean tree weight $\langle N_{\rm tree}\rangle$ (Eq.~\eqref{eq:L-Nrho}), and determine the exponent $\rho$ by using the methods of Sec.~\ref{sec:Theory-2dimBPs-Obs-Methods}.
The results for $\langle L\rangle$ {\it vs.} $\langle N_{\rm tree}\rangle$ for different $\kappa_{\rm bend}$ parameters are shown in panel (a) of Fig.~\ref{fig:SummarizingObs} (symbols) together with the estimated values and statistical errors for $\rho$ (dashed lines).
In particular, the behavior is in very good agreement with the results for $2d$ tree melts~\cite{RosaEveraersJCP2016} (Table~\ref{tab:ExpSummary-Obs}).
Interestingly, by introducing the asymptotic ({\it i.e.}, large-$\langle N_{\rm tree}\rangle$) fraction of branch-nodes $\Lambda$ (see r.h.s. panel in Fig.~\ref{fig:TreeMassVSRingMass} in SM~\cite{SMnote}), its inverse $\Lambda^{-1}$ corresponds to the mean path length between two branch-nodes~\cite{DaoudJoanny1981} and, so, to the natural length scale beyond which branching becomes effectively relevant.
As a consequence, data for individual $\kappa_{\rm bend}$ collapse onto a single master curve (Fig.~\ref{fig:SummarizingObs}(a), inset) by rescaling the $x$- and $y$-coordinate as the following:
\begin{eqnarray}
\langle N_{\rm tree} \rangle & \to & \frac{\langle N_{\rm tree} \rangle}{1/\Lambda} \, , \label{eq:Ntree-x-rescale} \\
\langle L \rangle & \to & \frac{\langle L \rangle}{1/\Lambda} \, \left( \frac{\langle N_{\rm tree} \rangle}{1/\Lambda} \right)^{\!\!-\rho} \, , \label{eq:Ntree-y-rescale}
\end{eqnarray}
where $\rho = 0.61$ in Eq.~\eqref{eq:Ntree-y-rescale} is the {\it average}\footnote{The excellent collapse in the inset of Fig.~\ref{fig:SummarizingObs}(a)  and the agreement between the measured $\rho$-values for individual $\kappa_{\rm bend}$ justifies {\it a posteriori} our choice for taking the average. In the rest of this work, we do the same also for the other exponents.} of the estimated best values for individual $\kappa_{\rm bend}$ (Table~\ref{tab:ExpSummary-Obs}).

Then, we consider the mean branch weight $\langle N_{\rm br}\rangle$ as a function of $\langle N_{\rm tree}\rangle$ (Eq.~\eqref{eq:Nbr-Neps}), and determine the related exponent $\epsilon$, see panel (b) in Fig.~\ref{fig:SummarizingObs} (symbols and dashed lines).
As before, our estimates for $\epsilon$ for individual $\kappa_{\rm bend}$ are in agreement with each other and with the corresponding measured values for $2d$ tree melts~\cite{RosaEveraersJCP2016} (Table~\ref{tab:ExpSummary-Obs}); moreover, it is important to emphasize that our estimated values satisfy also the relation $\epsilon=\rho$ with great accuracy.
Finally, data for individual $\kappa_{\rm bend}$ collapse onto a single master curve (Fig.~\ref{fig:SummarizingObs}(b), inset) by rescaling the $x$-coordinate according to Eq.~\eqref{eq:Ntree-x-rescale} and, analogously to Eq.~\eqref{eq:Ntree-y-rescale}, the $y$-coordinate as the following:
\begin{equation}\label{eq:Nbr-y-rescale}
\langle N_{\rm br} \rangle \to \frac{\langle N_{\rm br} \rangle}{1/\Lambda} \, \left( \frac{\langle N_{\rm tree} \rangle}{1/\Lambda} \right)^{\!\!-\epsilon} \, ,
\end{equation}
where $\epsilon = 0.63$ in Eq.~\eqref{eq:Nbr-y-rescale} is the average of the estimated best values for individual $\kappa_{\rm bend}$ (Table~\ref{tab:ExpSummary-Obs}).

Third, we consider the mean-square end-to-end spatial distance $\langle R_{\rm path}^2 \rangle$ as a function of the mean path length $\langle L\rangle$ (Eq.~\eqref{eq:R2ell}), and determine the related exponent $\nu_{\rm path}$, see panel (c) in Fig.~\ref{fig:SummarizingObs} (symbols and dashed lines).
Similarly to previous results, the estimated $\nu_{\rm path}$ for $\kappa_{\rm bend}=0$ and $\kappa_{\rm bend}=1$ agree well with each other and with the measured value for $2d$ tree melt (Table~\ref{tab:ExpSummary-Obs}); instead, the estimated value for $\kappa_{\rm bend}=1.5$ now is $\approx 13\%$ smaller.
Although this discrepancy can be attributed to the limited amount of branching observed for $\kappa_{\rm bend}=1.5$ compared to the structures in the other two cases (r.h.s. panel in Fig.~\ref{fig:TreeMassVSRingMass} in SM~\cite{SMnote}), the difference remains nonetheless small so we decide to ignore it and proceed with the analysis in the same manner as we did for the other observables.
In this regard, we notice that data for individual $\kappa_{\rm bend}$ are already on top of each other.
Nonetheless, it is useful to rescale the $x$- and $y$-coordinates as the following:
\begin{eqnarray}
\langle L \rangle & \to & \frac{\langle L \rangle}{1/\Lambda} \, , \label{eq:<L>-x-rescale} \\
\langle R_{\rm path}^2 \rangle & \to & \langle R_{\rm path}^2 \rangle \, \langle L \rangle^{-2\nu_{\rm path}} \, , \label{eq:<L>-y-rescale}
\end{eqnarray}
where $\nu_{\rm path} = 0.75$ in Eq.~\eqref{eq:<L>-y-rescale} is the average of the estimated best values for individual $\kappa_{\rm bend}$ (Table~\ref{tab:ExpSummary-Obs}).
The result (see inset of Fig.~\ref{fig:SummarizingObs}(c)) confirms the universal behavior and suggests that the mean-square end-to-end distance between two branch-nodes is $\simeq (1/\Lambda)^{2\nu_{\rm path}}$.

To conclude, we examine the mean-square gyration radius $\langle R_g^2 \rangle$ as a function of $\langle N_{\rm tree}\rangle$ (Eq.~\eqref{eq:Rg2-N2nu}), and determine the related exponent $\nu$, see panel (d) in Fig.~\ref{fig:SummarizingObs} (symbols and dashed lines).
Also in this last case, the estimated $\nu$ for $\kappa_{\rm bend}=0$ and $\kappa_{\rm bend}=1$ agree well with each other and with the measured value for $2d$ tree melt (Table~\ref{tab:ExpSummary-Obs}); again the estimated value for $\kappa_{\rm bend}=1.5$ is slightly ($\approx 8\%$) smaller, but we disregard this small difference in the analysis.
Overall, we confirm the general scaling relation $\nu=\rho\nu_{\rm path}$ relating the different exponents.
Finally, based on the results for the previous observables, we expect data for individual $\kappa_{\rm bend}$ to collapse on top of each other assuming Eq.~\eqref{eq:Ntree-x-rescale} for the rescaling of the $x$-coordinate and
\begin{equation}\label{eq:Rg2-y-rescale}
\langle R_g^2 \rangle \to \frac{\langle R_g^2 \rangle}{(1/\Lambda)^{2\nu_{\rm path}}} \, \left( \frac{\langle N_{\rm tree} \rangle}{1/\Lambda} \right)^{\!\!-2\nu}
\end{equation}
for the $y$-coordinate, and where $\nu = 0.49$ in Eq.~\eqref{eq:Rg2-y-rescale} is the average of the estimated best values for individual $\kappa_{\rm bend}$ (Table~\ref{tab:ExpSummary-Obs}).
The plot in the inset of Fig.~\ref{fig:SummarizingObs}(d) confirms our scaling ansatz.

To summarize this first part, the analysis of the primitive shapes of ring polymers in $2d$ melt by means of the same fundamental observables used to study melts of randomly branching polymers has shown that these latter systems are fundamentally analogous to the former.
In the next Section, we generalize the analysis to the distribution functions.

\subsection{Distribution functions}\label{sec:Results-PDFs}

%
\begin{figure}
\includegraphics[width=0.45\textwidth]{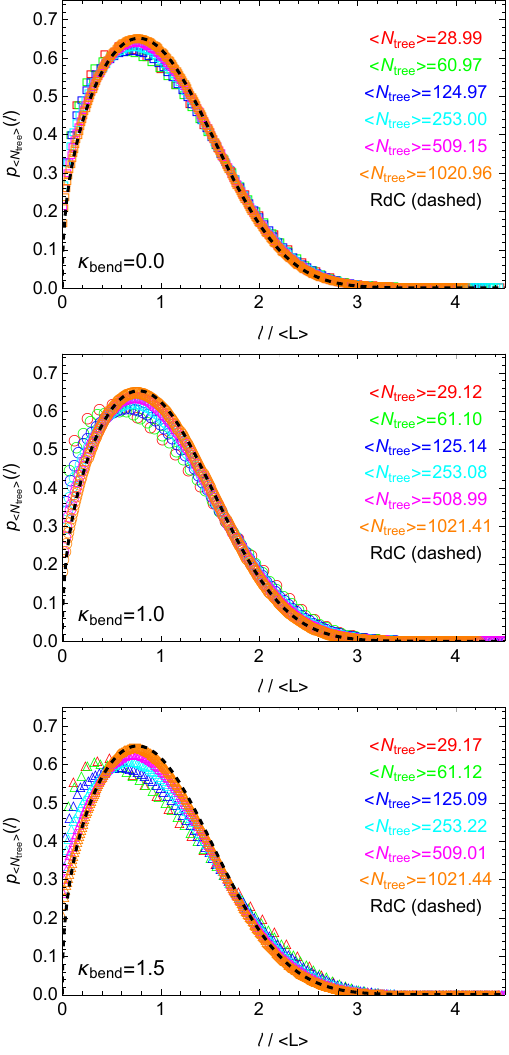}
\caption{
Conformational properties of trees: $p_{\langle N_{\rm tree}\rangle}(\ell)$, distribution functions of linear paths of length $\ell$.
The dashed line of each panel corresponds to the predicted RdC functional form for trees, Eqs.~\eqref{eq:q_RdC_path}-\eqref{eq:RdC_K}, with the exponents $\theta_{\ell}$ and $t_{\ell}$ as in Table~\ref{tab:ExpSummary-PDFs}.
Data of different colors denote different mean tree weight $\langle N_{\rm tree}\rangle$ (see legend).
}
\label{fig:PathLengthPDFs}
\end{figure}

We start with the path length distribution functions, $p_{\langle N_{\rm tree}\rangle}(\ell)$.
As illustrated in Fig.~\ref{fig:PathLengthPDFs}, the measured $p_{\langle N_{\rm tree}\rangle}(\ell)$ for different tree sizes fall onto universal master curves, when plotted as a function of the rescaled path length $x = \ell / \langle L \rangle$ (Eq.~\eqref{eq:pNell}), which are well described by the one-dimensional RdC form Eqs.~\eqref{eq:q_RdC_path}-\eqref{eq:RdC_K_l} (dashed lines).
Noticeably, the estimated values of the asymptotic exponents $(\theta_{\ell}, t_{\ell})$ (see Sec.~\ref{sec:Theory-2dimBPs-PDFs-Methods} for details) for the different chain flexibilities $\kappa_{\rm bend}$ are in good agreement with each other and are close to the measured values~\cite{RosaEveraers-TreesPDF2017} for trees in $2d$ melts (see Table~\ref{tab:ExpSummary-PDFs}).

\begin{figure}
\includegraphics[width=0.45\textwidth]{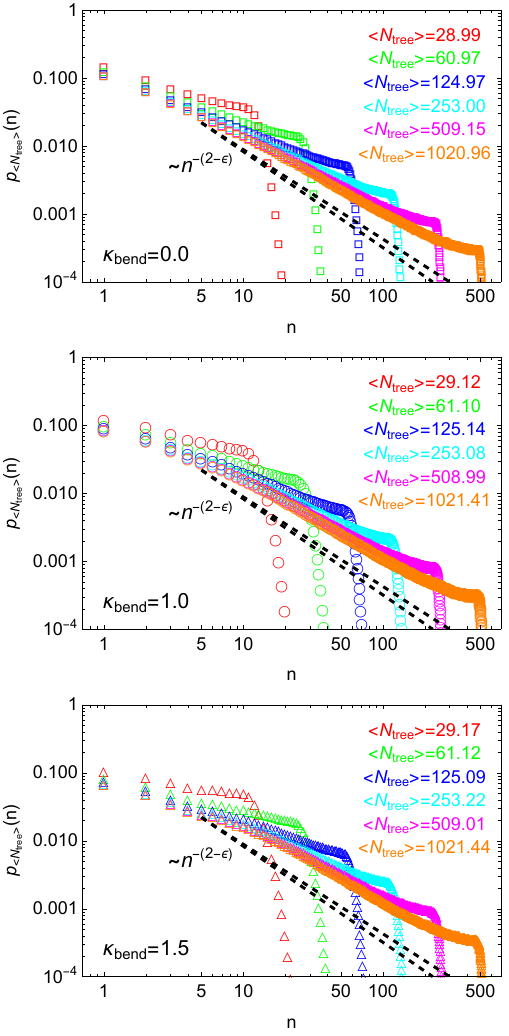}
\caption{
Conformational properties of trees: $p_{\langle N_{\rm tree}\rangle}(n)$, distribution functions of branch weight $n$.
The dashed lines express the interval of possible values for the exponent $\epsilon$ (see Fig.~\ref{fig:SummarizingObs}(b)), lying between the minimum lower-bound and the maximum upper-bound for all $\kappa_{\rm bend}$ (see Table~\ref{tab:ExpSummary-Obs}).
Colorcode/symbols are as in Fig.~\ref{fig:PathLengthPDFs}.
}
\label{fig:bweightPDFs}
\end{figure}

Next, we consider the distribution functions of branch weight $n$, $p_{\langle N_{\rm tree}\rangle}(n)$.
Data for different tree sizes $\langle N_{\rm tree}\rangle$ are in very good agreement (Fig.~\ref{fig:bweightPDFs}) with the predicted (Eq.~\eqref{eq:pNn}) power-law behavior $p_{\langle N_{\rm tree}\rangle}(n) \sim n^{-(2-\epsilon)}$ for trees (dashed lines, $\epsilon$ is as in panel (b) of Fig.~\ref{fig:SummarizingObs}).

\begin{figure}
\includegraphics[width=0.45\textwidth]{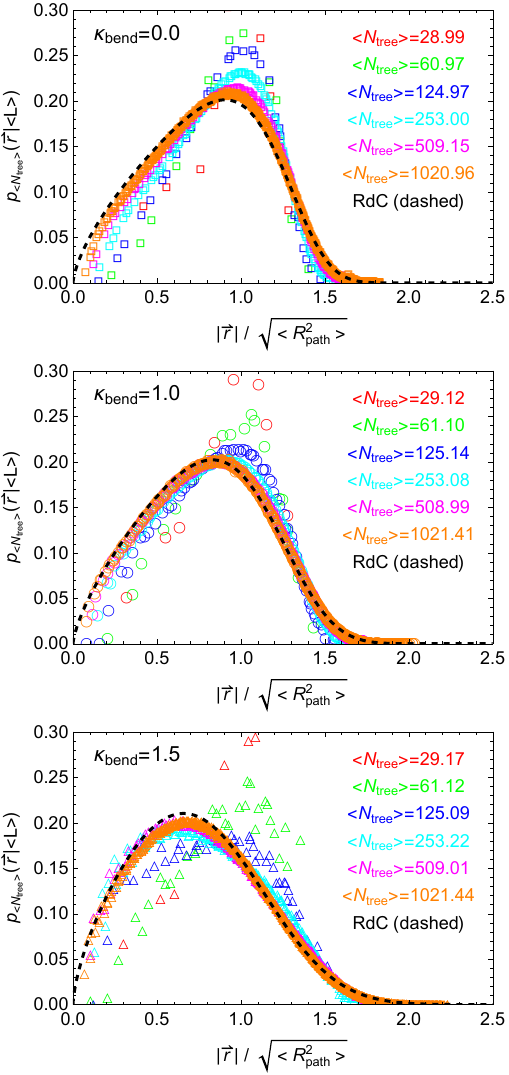}
\caption{
Conformational properties of trees: $p_{\langle N_{\rm tree}\rangle}({\vec r} | \langle L\rangle)$, distribution functions of end-to-end spatial distances $\vec r$ of paths of length $=\langle L\rangle$.
The dashed line of each panel corresponds to the RdC functional form for trees, Eq.~\eqref{eq:q_RdC_e2e_ell} with Eqs.~\eqref{eq:RdC_C}-\eqref{eq:RdC_K}, with the exponents $\theta_{\rm path}$ and $t_{\rm path}$ as in Table~\ref{tab:ExpSummary-PDFs}.
Colorcode/symbols are as in Fig.~\ref{fig:PathLengthPDFs}.
}
\label{fig:rijLPDFs}
\end{figure}

Then, we consider the distribution functions $p_{\langle N_{\rm tree}\rangle}({\vec r} | \langle L\rangle)$ of end-to-end spatial distances $\vec r$ of linear paths of length $=\langle L\rangle$.
When plotted as a function of the rescaled distance, $|\vec r| / \sqrt{\langle R_{\rm path}^2\rangle}$, the measured curves for different $\kappa_{\rm bend}$ fall onto single universal master curves (Fig.~\ref{fig:rijLPDFs}).
Here, however, and contrarily to the results presented so far, the master curves for the different stiffnesses agree less well with each other.
This, in particular, can be seen by the fact that the asymptotic RdC ansatz (dashed lines) are characterized by quite distinct values for the exponents $\theta_{\rm path}$ and, especially, $t_{\rm path}$ (see Table~\ref{tab:ExpSummary-PDFs}). 

\begin{figure}
\includegraphics[width=0.45\textwidth]{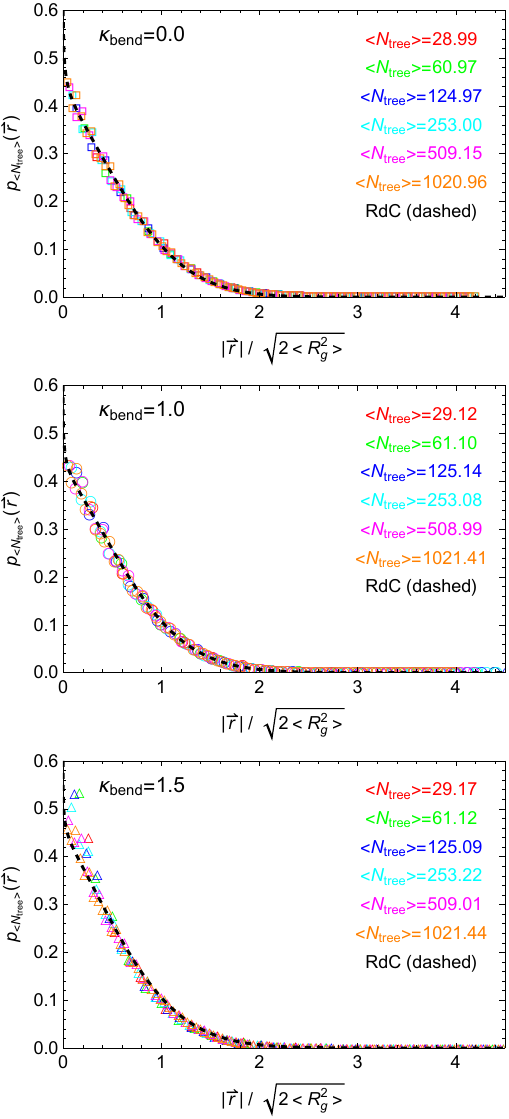}
\caption{
Conformational properties of trees: $p_{\langle N_{\rm tree}\rangle}({\vec r})$, distribution functions of vector distances $\vec r$ between tree nodes.
The dashed line of each panel corresponds to the RdC functional form for trees, Eq.~\eqref{eq:q_RdC_tree} with Eqs.~\eqref{eq:RdC_C}-\eqref{eq:RdC_K}, with the exponents $\theta_{\rm tree}$ and $t_{\rm tree}$ as in Table~\ref{tab:ExpSummary-PDFs}.
Colorcode/symbols are as in Fig.~\ref{fig:PathLengthPDFs}.
}
\label{fig:rijPDFs}
\end{figure}

Finally, we consider the distribution functions $p_{\langle N_{\rm tree}\rangle}({\vec r})$ of vector distances $\vec r$ between tree nodes.
By rescaling the $x$-axis by the square-root of the corresponding momenta $=\sqrt{2\langle R_g^2\rangle}$ curves from different rings obey universal behavior (Fig.~\ref{fig:rijPDFs}).
This behavior is well described (dashed lines) by the RdC function Eq.~\eqref{eq:q_RdC_tree}, with estimated values for the exponents $\theta_{\rm tree}$ and $t_{\rm tree}$ for the different flexibilities $\kappa_{\rm bend}$ in good agreement with each other and close to the estimated values for $2d$ tree melts (Table~\ref{tab:ExpSummary-PDFs}).

In conclusion, the results for both observables and distribution functions which have been presented so far demonstrate (besides some discrepancies detected in the distribution functions $p_{\langle N_{\rm tree}\rangle}({\vec r} | \langle L\rangle)$ for node-to-node spatial distances of linear paths) that the tree-like primitive paths of ring polymers in $2d$ melts obey the same statistics of $2d$ melts of randomly branching polymers.
We now explore the connection between these two systems in relation to polymer dynamics.

\section{A note on ring dynamics}\label{sec:Results-Dynamics}
Although it is not the main purpose of this work, we present here a few results on ring dynamics as well.
In fact, as originally pointed out by Rubinstein and coworkers~\cite{Rubinstein1986,Rubinstein1994} and reaffirmed at various degrees of sophistications in later works~\cite{SmrekGrosbergRingsR015,ElhamPRE2021,PanyukovRubinsteinMacromolecules2016}, fingerprints of the tree-like structure of rings in melt can be found in ring dynamics.

In particular, in Ref.~\cite{PanyukovRubinsteinMacromolecules2016} it was suggested that the relaxation time of the chains as a function of the monomers number $N_{\rm ring}$ is given by the power-law behavior
\begin{equation}\label{eq:TauRelaxRubinstein}
\tau_{\rm relax} = \tau_{\rm micr} \, N_{\rm ring}^{2+(1-\theta)\rho + \theta\nu} \, ,
\end{equation}
where $\tau_{\rm micr}$ is a ``microscopic'' time scale of the order of the time needed for the chain to relax between pairs of neighboring topological obstacles.
The $\rho$ and $\nu$ exponents in Eq.~\eqref{eq:TauRelaxRubinstein} are the same describing the behaviors of observables $\langle L \rangle$ (Eq.~\eqref{eq:L-Nrho}) and $\langle R_g^2\rangle$ (Eq.~\eqref{eq:Rg2-N2nu}) as functions of $N_{\rm ring}$, while the parameter $0\leq \theta \leq 1$ (not to be confused with the various exponents $\theta$ associated to the RdC functions) quantifies the postulated phenomenon of so called {\it tube dilation}: in analogy to linear chains~\cite{DoiEdwardsBook,RubinsteinColbyBook}, the array of topological obstacles where ring polymers relax can be assimilated to the relaxation inside a tube-like region shaped by the same obstacles.
Assuming such obstacles as fixed~\cite{SmrekGrosbergRingsR015}, the tube is a static structure and $\theta=0$; viceversa~\cite{PanyukovRubinsteinMacromolecules2016}, assuming that chain relaxation proceeds self-similarly at all scales (as in ordinary Rouse model~\cite{DoiEdwardsBook}), the tube ``dilate'' with time and $\theta=1$.
As shown in Ref.~\cite{PanyukovRubinsteinMacromolecules2016}, Eq.~\eqref{eq:TauRelaxRubinstein} applies well to the case of melt of rings in $3d$ with $\theta =1$, {\it i.e.} in good agreement with the mechanism of complete tube dilation.
What about the case of $2d$ melts, which we should expect to be {\it very} different from the $3d$ case because different chains can not, for instance, thread each other?

\begin{figure}
\includegraphics[width=0.45\textwidth]{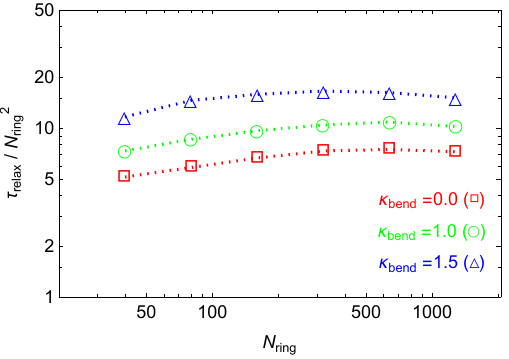}
\includegraphics[width=0.45\textwidth]{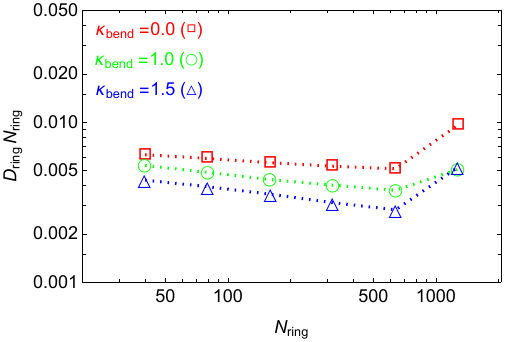}
\caption{
Ring dynamics:
(Top)
Ring relaxation time normalized by the square ring monomer number, $\tau_{\rm relax} / N_{\rm ring}^2$, as a function of $N_{\rm ring}$.
(Bottom)
Ring diffusion coefficient normalized by the inverse ring monomer number, $D_{\rm ring} / N_{\rm ring}^{-1}$, as a function of $N_{\rm ring}$.
Colorcode/symbols are for different chain stiffness $\kappa_{\rm bend}$ (see legend, as in Fig.~\ref{fig:SummarizingObs}).
Dotted lines joining the symbols serve as a guide for the eye.
Upraising of the data for $N_{\rm ring}=1280$ is due to the limited length of the MC trajectory (see Fig.~\ref{fig:g3Plot} in SM~\cite{SMnote}).
}
\label{fig:tauRelaxVSNring}
\end{figure}

To answer this question, we have computed $\tau_{\rm relax}$ for our rings by defining it as the time scale for the entire ring to diffuse by its size, {\it i.e.}
\begin{equation}\label{eq:DefineTauRelax}
g_3(t=\tau_{\rm relax}) = \langle R_g^2\rangle_{\rm ring} \, ,
\end{equation}
where
\begin{equation}\label{eq:Introduce-g3}
g_3(t) \equiv \left\langle ( \vec r_{\rm cm}(t) - \vec r_{\rm cm}(0) )^2 \right\rangle
\end{equation}
is the mean-square displacement of the chain centre of mass.
Results for $\tau_{\rm relax}$, normalized by $N_{\rm ring}^2$, as a function of $N_{\rm ring}$ are shown in the top panel of Fig.~\ref{fig:tauRelaxVSNring}.
Interestingly the data suggest the scaling $\tau_{\rm relax} \sim N_{\rm ring}^2$, therefore they are not compatible with the power-law~\eqref{eq:TauRelaxRubinstein} whose exponent is, by construction, strictly $>2$.
To validate this result further, we consider the scaling behavior of the chain diffusion coefficient,
\begin{equation}\label{eq:Dring}
D_{\rm ring} \equiv \lim_{t\to\infty} \frac{g_3(t)}{4t} \, ,
\end{equation}
for which we expect $D_{\rm ring} \sim \langle R_g^2\rangle / \tau_{\rm ring} \sim N_{\rm ring}^{-2(1-\nu)} = N_{\rm ring}^{-1}$.
We have estimated diffusion coefficients by best fits of data for $g_3(t)$ to a simple linear function (Eq.~\eqref{eq:Dring}), with the fit procedure restricted to the time interval $t>5\tau_{\rm relax}$ (see Fig.~\ref{fig:g3Plot} in SM~\cite{SMnote}).
Results for $D_{\rm ring}$, each multiplied by the corresponding $N_{\rm ring}$, are shown in the bottom panel of Fig.~\ref{fig:tauRelaxVSNring} for each stiffness $\kappa_{\rm bend}$, and they confirm the scaling analysis.

Curiously, we notice that the power-law behaviors $\tau_{\rm relax} \sim N_{\rm ring}^2$ and $D_{\rm ring} \sim N_{\rm ring}^{-1}$ reported here are the same that one would have expected for an ordinary Rouse chain in ideal ({\it i.e.}, no volume interactions) conditions~\cite{DoiEdwardsBook,RubinsteinColbyBook}.
For the polymers considered here this is obviously not the case, and the agreement with the Rouse model must be considered a fortuituous coincidence for which, at the moment, we have no physical intuition.
We just conclude by highlighting the fact that the two power-laws are no artefact of the present lattice model, as they were also reported in previous molecular simulations of $2d$ ring melts~\cite{KimBaigSoftMatter2021}.

\section{Summary and conclusions}\label{sec:Concls}
In this work we have performed a detailed analysis of the conformational properties of unknotted and non-concatenated ring polymers in $2d$ melt conditions (Fig.~\ref{fig:Primitive-Path}), aiming at demonstrating that rings double-fold around randomly branching (tree-like) conformations as originally proposed by Khokhlov, Nechaev, Rubinstein and others~\cite{NechaevKhokhlov1985,NechaevKhokhlov1985,Rubinstein1986,Rubinstein1994}.
In particular, after reconstructing the ``primitive shapes'' of ring conformations from Monte Carlo computer simulations of a generic polymer model on the two-dimensional triangular lattice, we proceeded with an analysis of said branched structures by means of the same observables and distribution functions (Sec.~\ref{sec:Theory-2dimBPs}) used for the analysis of lattice trees.
We thus confirm (Sec.~\ref{sec:Results}, in particular Figs.~\ref{fig:SummarizingObs}-\ref{fig:rijPDFs}) that the conformational properties of the branched structures are in very good agreement with the measured properties~\cite{RosaEveraersJCP2016,RosaEveraers-TreesPDF2017} of two-dimensional branched polymers in melt conditions.

Quite interestingly, the analysis of ring dynamics (Sec.~\ref{sec:Results-Dynamics}) suggests instead a rather different behavior than the one expected based on the branched shapes of the rings.
In fact, the rings' relaxation time is found to be {\it faster} than the theoretical predictions, scaling ``only'' quadratically with the rings' mass and, therefore, surprisingly close to the result expected for ordinary Rouse motion.

We conclude with outlook for future work.
The next natural step should focus on the study of {\it three}-dimensional melts of rings.
As a matter of fact, how rings fold in $3d$ is still not fully understood.
In fact, while on one side a multi-scale numerical protocol based on the random-tree model~\cite{RosaEveraersPRL2014} reveals an accurate similarity between some of the trees' and rings' static properties, on the other the ``tree picture'' cannot be considered complete since, for example, one of two nearby rings may easily penetrate onto an open loop of the other without violation of topological constraints.
Although rare, such so-called {\it threading} events~\cite{MichielettoSM2016,SmrekGrosbergACSMacroLett2016,SmrekRosa2019,SchramRosaEveraers2019,UbertiniSmrekRosa2022} have been observed and are thought to be responsible of some exotic properties in rings' out-of-equilibrium behavior~\cite{Michieletto2016,VlassopoulosPRL2019,OConnorPRL2020,SmrekChubak2020,MichelettiSmrek2024}.
For these reasons, developing an algorithm for the reconstruction of the chain's ``primitive shape'' along the lines of Sec.~\ref{sec:RingsPrimitiveShape} would represent an important contribution towards the comprehension of these fundamental systems.

{\it Conflicts of interest} --
There are no conflicts to declare.

{\it Data availability} --
The data supporting this article have been included as part of the Supplemental Material.

{\it Acknowledgements} --
MAU is supported by an EMBO Postdoctoral Fellowship (ALTF 1305-2024).
AR acknowledges S.K. Nechaev for insightful discussions, and financial support from PNRR Grant CN\_00000013\_CN-HPC, M4C2I1.4, spoke 7, funded by Next Generation EU.


\bibliography{../biblio}

\clearpage

\widetext
\clearpage
\begin{center}
\textbf{\Large Supplemental Material \\ \vspace*{1.5mm} Ring polymers in two-dimensional melts double-fold around randomly branching ``primitive shapes''} \\
\vspace*{5mm}
Mattia Alberto Ubertini, Angelo Rosa
\vspace*{10mm}
\end{center}

\setcounter{equation}{0}
\setcounter{figure}{0}
\setcounter{table}{0}
\setcounter{page}{1}
\setcounter{section}{0}
\setcounter{page}{1}
\makeatletter
\renewcommand{\theequation}{S\arabic{equation}}
\renewcommand{\thefigure}{S\arabic{figure}}
\renewcommand{\thetable}{S\arabic{table}}
\renewcommand{\thesection}{S\arabic{section}}
\renewcommand{\thepage}{S\arabic{page}}

\tableofcontents

\clearpage
\section*{Supplemental tables}

\clearpage
\begin{table*}
\begin{tabular}{cccccc}
$N_{\rm ring}$ & $\langle N_{\rm tree}\rangle$ & $\langle L\rangle \, [a/\sqrt{3}]$ & $\langle N_{\rm br}\rangle$ & $\langle R_{\rm path}^2\rangle \, [a^2]$ & $\langle R_g^2\rangle \, [a^2]$ \\
\hline
\hline
\\
\\
\multicolumn{6}{c}{$\kappa_{\rm bend}=0$} \\
\hline
\\
$40$ & $29.0 \pm 3.4$ & $6.954 \pm 0.002$ & $3.730 \pm 0.001$ & $7.446 \pm 0.002$ & $4.220 \pm 0.002$ \\
$80$ & $61.0 \pm 5.1$ & $12.131 \pm 0.004$ & $7.301 \pm 0.004$ & $18.264 \pm 0.006$ & $10.369 \pm 0.008$ \\
$160$ & $125.0 \pm 7.4$ & $20.414 \pm 0.009$ & $12.931 \pm 0.009$ & $41.61 \pm 0.02$ & $23.37 \pm 0.03$ \\
$320$ & $253.0 \pm 10.6$ & $33.47 \pm 0.02$ & $21.72 \pm 0.02$ & $91.70 \pm 0.07$ & $50.15 \pm 0.08$ \\
$640$ & $509.1 \pm 15.2$ & $53.78 \pm 0.05$ & $35.31 \pm 0.05$ & $198.2 \pm 0.2$ & $104.6 \pm 0.2$ \\
$1280$ & $1021.0 \pm 21.2$ & $85.4 \pm 0.1$ & $56.4 \pm 0.1$ & $402.5 \pm 0.7$ & $214.5 \pm 0.7$ \\
\hline
\\
\\
\multicolumn{6}{c}{$\kappa_{\rm bend}=1$} \\
\hline
\\
$40$ & $29.1 \pm 4.1$ & $7.792 \pm 0.002$ & $4.406 \pm 0.002$ & $9.783 \pm 0.003$ & $5.384 \pm 0.004$ \\
$80$ & $61.1 \pm 6.1$ & $13.727 \pm 0.005$ & $8.487 \pm 0.005$ & $23.74 \pm 0.01$ & $12.49 \pm 0.01$ \\
$160$ & $125.1 \pm 8.8$ & $23.01 \pm 0.01$ & $14.75 \pm 0.01$ & $50.33 \pm 0.03$ & $26.43 \pm 0.04$ \\
$320$ & $253.1 \pm 12.5$ & $37.45 \pm 0.03$ & $24.41 \pm 0.02$ & $102.6 \pm 0.1$ & $54.20 \pm 0.1$ \\
$640$ & $509.0 \pm 17.6$ & $59.81 \pm 0.06$ & $39.31 \pm 0.06$ & $212.2 \pm 0.3$ & $109.8 \pm 0.3$ \\
$1280$ & $1021.4 \pm 25.1$ & $94.0 \pm 0.1$ & $62.0 \pm 0.1$ & $422.8 \pm 0.9$ & $219.6 \pm 0.7$ \\
\hline
\\
\\
\multicolumn{6}{c}{$\kappa_{\rm bend}=1.5$} \\
\hline
\\
$40$ & $29.2 \pm 4.6$ & $8.333 \pm 0.002$ & $4.880 \pm 0.002$ & $10.941 \pm 0.005$ & $7.224 \pm 0.006$ \\
$80$ & $61.1 \pm 6.8$ & $15.257 \pm 0.006$ & $9.739 \pm 0.005$ & $30.61 \pm 0.02$ & $17.65 \pm 0.03$ \\
$160$ & $125.1 \pm 9.9$ & $26.05 \pm 0.01$ & $17.05 \pm 0.01$ & $67.91 \pm 0.07$ & $35.21 \pm 0.07$ \\
$320$ & $253.2 \pm 14.0$ & $42.74 \pm 0.03$ & $28.19 \pm 0.03$ & $135.1 \pm 0.2$ & $67.0 \pm 0.2$ \\
$640$ & $509.0 \pm 20.0$ & $68.24 \pm 0.08$ & $45.08 \pm 0.07$ & $253.4 \pm 0.5$ & $126.9 \pm 0.4$ \\
$1280$ & $1021.4 \pm 28.7$ & $107.0 \pm 0.2$ & $70.8 \pm 0.2$ & $482.2 \pm 1.4$ & $243.6 \pm 1.0$ \\
\hline
\end{tabular}
\caption{
Conformational properties of trees (see Sec.~\ref{sec:Theory-2dimBPs-Obs-Defs} in main text for details):
$\langle N_{\rm tree}\rangle$, mean tree weight;
$\langle L\rangle$, mean path length;
$\langle N_{\rm br}\rangle$, mean branch weight;
$\langle R_{\rm path}^2\rangle$, mean-square end-to-end spatial distance of paths of contour length $=\langle L\rangle$;
$\langle R_g^2\rangle$, mean-square gyration radius.
For reference, the first column reports the number of monomers, $N_{\rm ring}$, of the original rings.
$a$ is the unit of length, equal to the spacing of the original triangular lattice where rings were simulated (see Secs.~\ref{sec:PolymerModel} and~\ref{sec:RingsPrimitiveShape} in main text for details).
}
\label{tab:TreeObsMeanValues}
\end{table*}

\clearpage
\begin{table*}
\begin{tabular}{cccc}
\,\,\,\,\, $\langle L\rangle \sim \langle N_{\rm tree}\rangle^\rho$ \,\,\,\,\, & \,\,\,\,\, $\langle N_{\rm br}\rangle \sim \langle N_{\rm tree}\rangle^\epsilon$ \,\,\,\,\, & \,\,\,\,\, $\langle R_{\rm path}^2\rangle \sim \langle L\rangle^{2\nu_{\rm path}}$ \,\,\,\,\, & \,\,\,\,\, $\langle R_g^2\rangle \sim \langle N_{\rm tree}\rangle^{2\nu}$ \,\,\,\,\, \\
\hline
\hline
\\
\\
\multicolumn{4}{c}{$\kappa_{\rm bend}=0$, Eq.~\eqref{eq:Fit-2params} in main text} \\
\hline
$c=-0.204$ & $c=-0.706$ & $c=-1.018$ & $c=-1.847$ \\
$\rho=0.672$ & $\epsilon=0.684$ & $\nu_{\rm path}=0.790$ & $\nu=0.521$ \\
\\
\multicolumn{4}{c}{$\kappa_{\rm bend}=0$, Eq.~\eqref{eq:Fit-4params} in main text} \\
\hline
$c=1.736$ & $c=0.343$ & $c=1.417$ & $c=-0.600$ \\
$d=-2.741$ & $d=-5.065$ & $d=-2.507$ & $d=-3.657$ \\
$\Delta=0.349$ & $\Delta=1.331$ & $\Delta=0.051$ & $\Delta=0.963$ \\
$\rho=0.593$ & $\epsilon=0.588$ & $\nu_{\rm path}=0.777$ & $\nu=0.472$ \\
\hline
\\
\\
\multicolumn{4}{c}{$\kappa_{\rm bend}=1$, Eq.~\eqref{eq:Fit-2params} in main text} \\
\hline
$c=-0.026$ & $c=-0.500$ & $c=-0.941$ & $c=-1.554$ \\
$\rho=0.660$ & $\epsilon=0.684$ & $\nu_{\rm path}=0.769$ & $\nu=0.501$ \\
\\
\multicolumn{4}{c}{$\kappa_{\rm bend}=1$, Eq.~\eqref{eq:Fit-4params} in main text} \\
\hline
$c=2.420$ & $c=0.252$ & $c=-0.764$ & $c=-1.472$ \\
$d=-3.479$ & $d=-5.726$ & $d=-0.126$ & $d=-26.576$ \\
$\Delta=0.361$ & $\Delta=1.653$ & $\Delta=-0.900${}$^{(\ast)}$ & $\Delta=4.068$ \\
$\rho=0.556$ & $\epsilon=0.593$ & $\nu_{\rm path}=0.803$ & $\nu=0.496$ \\
\hline
\\
\\
\multicolumn{4}{c}{$\kappa_{\rm bend}=1.5$, Eq.~\eqref{eq:Fit-2params} in main text} \\
\hline
$c=0.116$ & $c=-0.311$ & $c=-0.305$ & $c=-0.917$ \\
$\rho=0.658$ & $\epsilon=0.660$ & $\nu_{\rm path}=0.693$ & $\nu=0.463$ \\
\\
\multicolumn{4}{c}{$\kappa_{\rm bend}=1.5$, Eq.~\eqref{eq:Fit-4params} in main text} \\
\hline
$c=2.934$ & $c=0.274$ & $c=-0.194$ & $c=-0.876$ \\
$d=-4.484$ & $d=-10.644$ & $d=-9.755$ & $d=-88132.300${}$^{(\ast)}$ \\
$\Delta=0.465$ & $\Delta=2.243$ & $\Delta=4.6118$ & $\Delta=10.519$ \\
$\rho=0.514$ & $\epsilon=0.595$ & $\nu_{\rm path}=0.682$ & $\nu=0.459$ \\
\hline
\end{tabular}
\caption{
Optimal parameters obtained by best fits of Eq.~\eqref{eq:Fit-2params} and Eq.~\eqref{eq:Fit-4params} in main text to the data for tree observables (see top row here, Table~\ref{tab:TreeObsMeanValues} and Sec.~\ref{sec:Theory-2dimBPs-Obs-Methods} in main text for details).
{}$^{(\ast)}$In these two cases, the four-parameter model does not provide completely accurate results: nonetheless, the values for the scaling exponents look reasonable and we decide to employ them.
}
\label{tab:TreeObsFitResults}
\end{table*}

\clearpage
\begin{table*}
\begin{tabular}{cccccccc}
$N_{\rm ring}$ & $\langle N_{\rm tree}\rangle$ & $\theta_{\ell}$ & $t_{\ell}$ & $\theta_{\rm path}$ & $t_{\rm path}$ & $\theta_{\rm tree}$ & $t_{\rm tree}$ \\
\hline
\hline
\\
\\
\multicolumn{8}{c}{$\kappa_{\rm bend}=0$} \\
\hline
\\
$40$ & $29.0 \pm 3.4$ & $0.25 \pm 0.02$ & $2.53 \pm 0.07$ & $2.7 \pm 0.7$ & $10.8 \pm 3.4$ & $-0.35 \pm 0.05$ & $2.41 \pm 0.09$ \\
$80$ & $61.0 \pm 5.1$ & $0.242 \pm 0.007$ & $2.65 \pm 0.04$ & $2.0 \pm 0.1$ & $10.2 \pm 0.8$ & $-0.27 \pm 0.03$ & $2.17 \pm 0.08$ \\
$160$ & $125.0 \pm 7.4$ & $0.266 \pm 0.002$ & $2.72 \pm 0.01$ & $1.58 \pm 0.03$ & $9.3 \pm 0.2$ & $-0.20 \pm 0.03$ & $2.00 \pm 0.07$ \\
$320$ & $253.0 \pm 10.6$ & $0.3066 \pm 0.0009$ & $2.731 \pm 0.005$ & $1.115 \pm 0.009$ & $8.72 \pm 0.08$ & $-0.16 \pm 0.02$ & $1.91 \pm 0.05$ \\
$640$ & $509.1 \pm 15.2$ & $0.353 \pm 0.001$ & $2.703 \pm 0.006$ & $0.887 \pm 0.006$ & $7.56 \pm 0.05$ & $-0.13 \pm 0.02$ & $1.84 \pm 0.04$ \\
\,\, $1280$ \,\, & \,\,\, $1021.0 \pm 21.2$ \,\,\, & \,\,\, $0.398 \pm 0.001$ \,\,\, & \,\,\, $2.645 \pm 0.005$ \,\,\, & \,\, $0.791 \pm 0.003$ \,\, & \,\,\, $6.70 \pm 0.02$ \,\,\, & \,\, $-0.11 \pm 0.01$ \,\, & \,\,$1.79 \pm 0.03$ \,\, \\
\hline
\\
\\
\multicolumn{8}{c}{$\kappa_{\rm bend}=1$} \\
\hline
\\
$40$ & $29.1 \pm 4.1$ & $0.15 \pm 0.02$ & $2.40 \pm 0.07$ & $1.8 \pm 0.2$ & $11.0 \pm 2.2$ & $-0.36 \pm 0.06$ & $1.92 \pm 0.09$ \\
$80$ & $61.1 \pm 6.1$ & $0.171 \pm 0.006$ & $2.54 \pm 0.03$ & $1.41 \pm 0.08$ & $8.8 \pm 0.6$ & $-0.28 \pm 0.04$ & $1.91 \pm 0.07$ \\
$160$ & $125.1 \pm 8.8$ & $0.221 \pm 0.002$ & $2.650 \pm 0.009$ & $0.96 \pm 0.03$ & $7.6 \pm 0.3$ & $-0.18 \pm 0.03$ & $1.83 \pm 0.06$ \\
$320$ & $253.1 \pm 12.5$ & $0.280 \pm 0.002$ & $2.69 \pm 0.01$ & $0.73 \pm 0.01$ & $6.37 \pm 0.07$ & $-0.13 \pm 0.02$ & $1.79 \pm 0.05$ \\
$640$ & $509.0 \pm 17.6$ & $0.343 \pm 0.002$ & $2.65 \pm 0.01$ & $0.716 \pm 0.007$ & $5.43 \pm 0.04$ & $-0.10 \pm 0.02$ & $1.76 \pm 0.04$ \\
$1280$ & $1021.4 \pm 25.1$ & $0.400 \pm 0.002$ & $2.624 \pm 0.008$ & $0.718 \pm 0.004$ & $5.13 \pm 0.02$ & $-0.08 \pm 0.01$ & $1.74 \pm 0.03$ \\
\hline
\\
\\
\multicolumn{8}{c}{$\kappa_{\rm bend}=1.5$} \\
\hline
\\
$40$ & $29.2 \pm 4.6$ & $0.09 \pm 0.02$ & $2.34 \pm 0.09$ & $1.7 \pm 0.1$ & $326.8 \pm 27649.7${}$^{(\ast)}$ & $-0.40 \pm 0.06$ & $1.50 \pm 0.06$ \\
$80$ & $61.1 \pm 6.8$ & $0.105 \pm 0.008$ & $2.39 \pm 0.04$ & $0.9 \pm 0.1$ & $11.8 \pm 2.1$ & $-0.29 \pm 0.04$ & $1.46 \pm 0.05$ \\
$160$ & $125.1 \pm 9.9$ & $0.154 \pm 0.002$ & $2.54 \pm 0.01$ & $0.47 \pm 0.06$ & $7.0 \pm 0.6$ & $-0.17 \pm 0.03$ & $1.47 \pm 0.04$ \\
$320$ & $253.2 \pm 14.0$ & $0.222 \pm 0.002$ & $2.62 \pm 0.01$ & $0.33\pm 0.02$ & $4.5 \pm 0.1$ & $-0.11 \pm 0.02$ & $1.52 \pm 0.03$ \\
$640$ & $509.0 \pm 20.0$ & $0.299 \pm 0.002$ & $2.62 \pm 0.01$ & $0.50 \pm 0.01$ & $3.48 \pm 0.04$ & $-0.09 \pm 0.02$ & $1.58 \pm 0.03$ \\
$1280$ & $1021.4 \pm 28.7$ & $0.375 \pm 0.002$ & $2.57 \pm 0.01$ & $0.576 \pm 0.004$ & $3.46 \pm 0.01$ & $-0.07 \pm 0.01$ & $1.62 \pm 0.02$ \\
\hline
\end{tabular}
\caption{
Estimated values for the pairs of exponents
($\theta_{\ell} \pm \Delta\theta_{\ell}$, $t_{\ell} \pm \Delta t_{\ell}$),
($\theta_{\rm path} \pm \Delta\theta_{\rm path}$, $t_{\rm path} \pm \Delta t_{\rm path}$),
($\theta_{\rm tree} \pm \Delta\theta_{\rm tree}$, $t_{\rm tree} \pm \Delta t_{\rm tree}$)
of, respectively, the Redner-des Cloizeaux (RdC) functions Eq.~\eqref{eq:q_RdC_path}, Eq.~\eqref{eq:q_RdC_e2e_ell} and Eq.~\eqref{eq:q_RdC_tree} in main text.
For each $\langle N_{\rm tree}\rangle$, each pair of exponents was calculated by best fit of the RdC function to the corresponding data set for $p_{\langle N_{\rm tree}\rangle}(\cdot)$ (see Fig.~\ref{fig:PathLengthPDFs}, Fig.~\ref{fig:rijLPDFs} and Fig.~\ref{fig:rijPDFs} in main text).
Additional details concerning the fit procedure are given in Sec.~\ref{sec:Theory-2dimBPs-PDFs-Methods} in main text.
{}$^{(\ast)}$This unrealistic value is of course due to the fact that the RdC curve is, in this case, not a good model for the measured distribution function.
}
\label{tab:TreePDFsFitResults}
\end{table*}

\clearpage
\section*{Supplemental figures}

\clearpage
\begin{figure}
\includegraphics[width=0.50\textwidth]{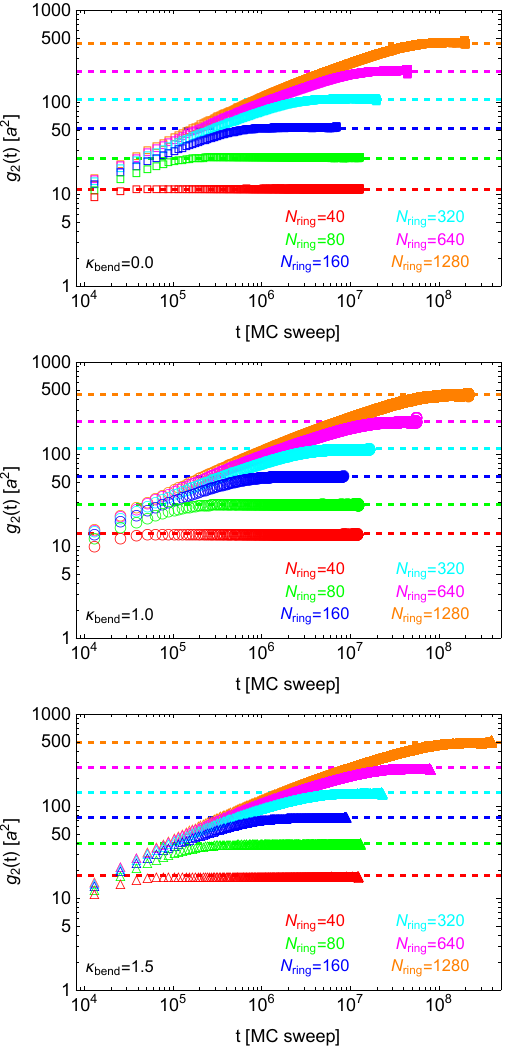}
\caption{
Ring dynamics:
$g_2(t)$, time mean-square displacement of monomers relative to the chain center of mass (see Sec.~\ref{sec:SimulationDetails} in main text for definitions, in particular Eq.~\eqref{eq:Introduce-g2}).
The three panels summarize all considered systems.
At large times, $g_2(t) \to 2\langle R_g^2\rangle_{\rm ring}$ where $\langle R_g^2\rangle_{\rm ring}$ is the ring mean-square gyration radius (dashed lines, see Eq.~\eqref{eq:<Rg2>_ring} in main text).
Colorcode/symbols are as in Fig.~\ref{fig:PathLengthPDFs} in main text, with the legend here reporting the number of monomers per ring $N_{\rm ring}$.
}
\label{fig:g2Plot}
\end{figure}
%

\clearpage
\begin{figure}
\includegraphics[width=0.5\textwidth]{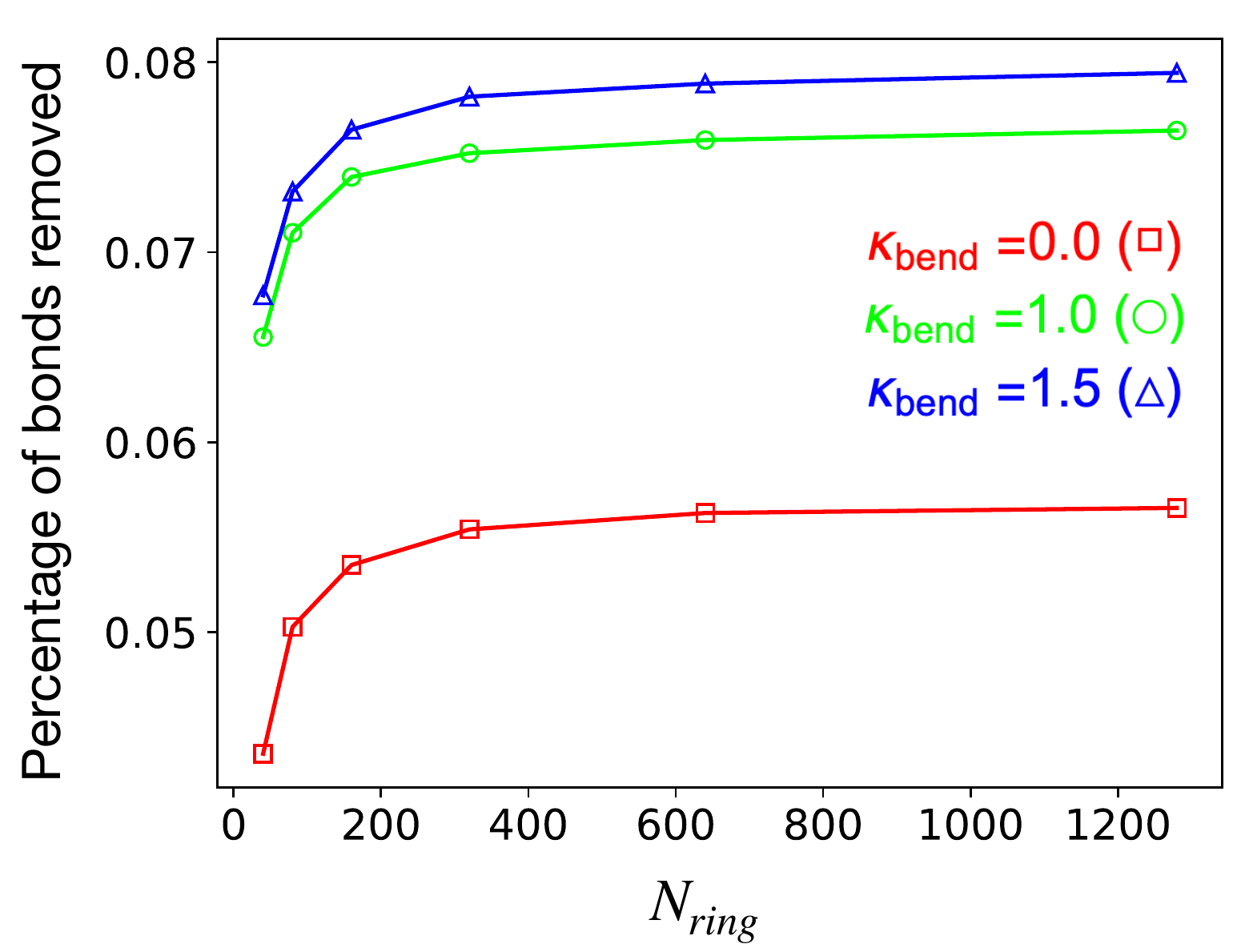}
\caption{
Percentage of bonds that were cut from the reconstructed trees in order to obtain final structures with no loops, plotted as a function of the polymer ring size $N_{\rm ring}$ from which the trees were derived.
Colorcode/symbols are as in Fig.~\ref{fig:SummarizingObs} in main text.
}
\label{fig:Statistics_Bond_Removed}
\end{figure}

\clearpage
\begin{figure}
$$
\begin{array}{cc}
\includegraphics[width=0.45\textwidth]{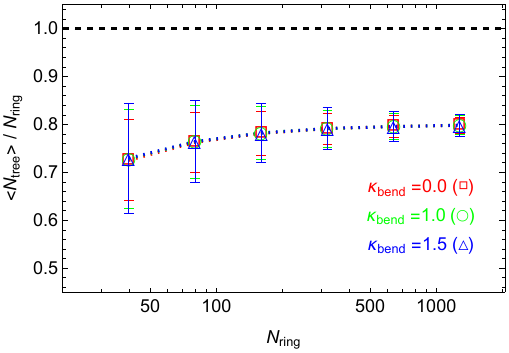} & \includegraphics[width=0.45\textwidth]{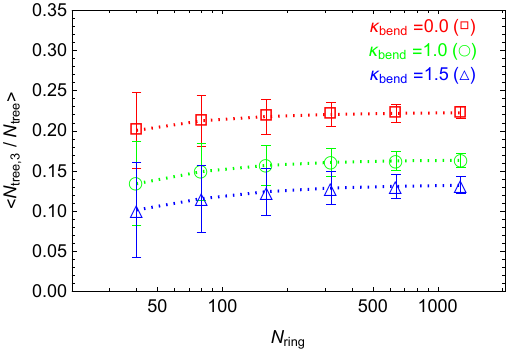}
\end{array}
$$
\caption{
(left)
Mean number of tree nodes normalized to ring mass, $\langle N_{\rm tree}\rangle / N_{\rm ring}$, as a function of $N_{\rm ring}$ and for ring bending stiffness $\kappa_{\rm bend}$.
The horizontal line $y=1$ (dashed) is shown for comparison.
(right)
Mean fraction of branch-nodes, $\langle N_{{\rm tree}, 3} / N_{\rm tree} \rangle$, as a function of $N_{\rm ring}$.
In both panels: the vertical bars on the symbols denote the measured standard deviations, while dotted lines joining the symbols serve as a guide for the eye.
Colorcode/symbols are as in Fig.~\ref{fig:SummarizingObs} in main text.
}
\label{fig:TreeMassVSRingMass}
\end{figure}

\clearpage
\begin{figure*}
$$
\begin{array}{cc}
\multicolumn{2}{c}{\rm (a)} \\
\includegraphics[width=0.45\textwidth]{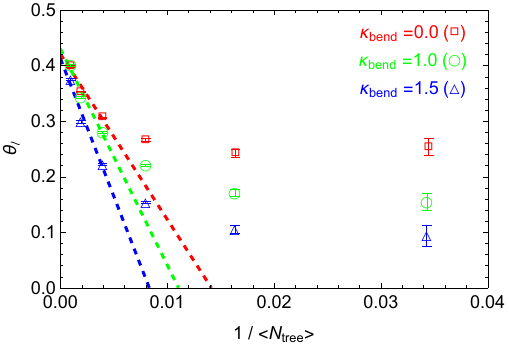} & \includegraphics[width=0.45\textwidth]{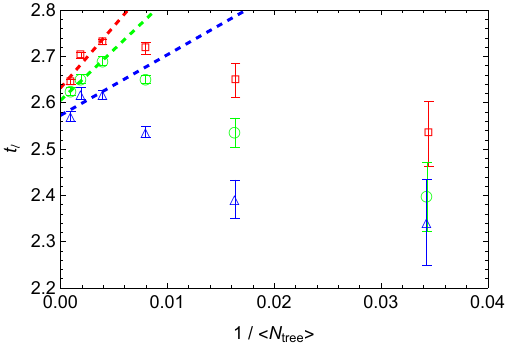} \\
\\
\multicolumn{2}{c}{\rm (b)} \\
\includegraphics[width=0.45\textwidth]{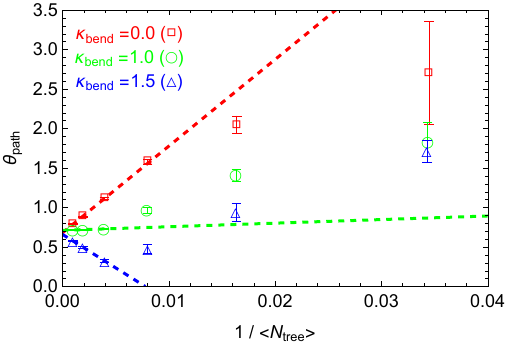} & \includegraphics[width=0.45\textwidth]{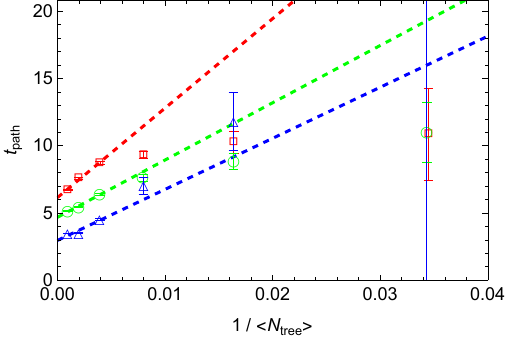}
\\
\multicolumn{2}{c}{\rm (c)} \\
\includegraphics[width=0.45\textwidth]{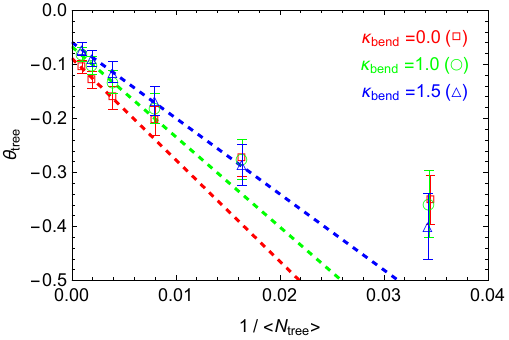} & \includegraphics[width=0.45\textwidth]{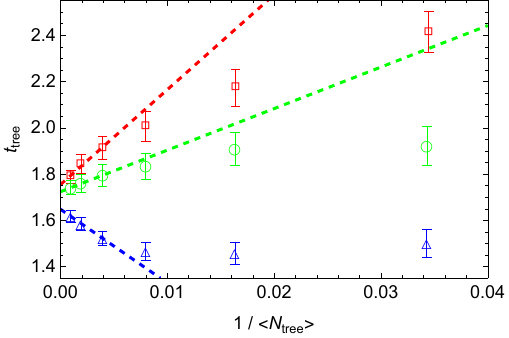}
\end{array}
$$
\caption{
Graphical illustration of the extrapolation procedure to get asymptotic ({\it i.e.}, for $\langle N_{\rm tree}\rangle \to\infty$) values of pairs of exponents
($\theta_{\ell}, \, t_{\ell}$),
($\theta_{\rm path}, \, t_{\rm path}$),
($\theta_{\rm tree}, \, t_{\rm tree}$).
Symbols are for the estimated values at finite $\langle N_{\rm tree}\rangle$ (see Table~\ref{tab:TreePDFsFitResults}), shown here as a function of $1/\langle N_{\rm tree}\rangle$.
Dashed lines are best fits of the expression $y = -Ax + B$ to the data for the three largest $\langle N_{\rm tree}\rangle$'s and with fit parameters $A$ and $B$.
The intercept with the $y$-axis, $B$, provides the estimate for the asymptotic value of the exponent.
Additional details on the procedure are given in Sec.~\ref{sec:Theory-2dimBPs-PDFs-Methods} in main text.
Colorcode/symbols are as in Fig.~\ref{fig:SummarizingObs} in main text.
}
\label{fig:RdC_estimatedparams_and_errors}
\end{figure*}
%

\clearpage
\begin{figure}
\includegraphics[width=0.50\textwidth]{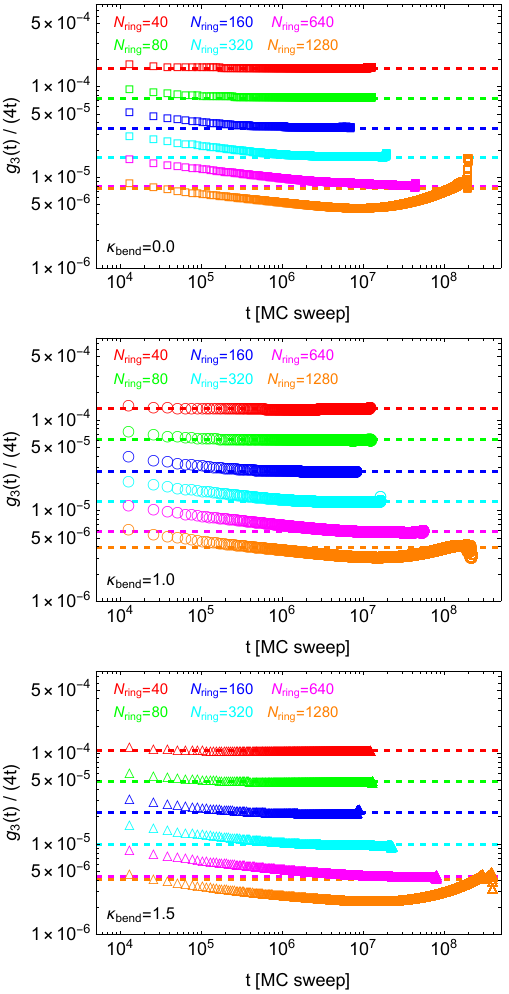}
\caption{
Ring dynamics:
$g_3(t)$, time mean-square displacement of the chain center of mass (see Eq.~\eqref{eq:Introduce-g3} in main text) normalized by $4t$.
The dashed lines are best fits of the data to a constant value in the interval $t>5\tau_{\rm relax}$, which is used to determine the chain diffusion coefficient, $D_{\rm ring}$ (see Eq.~\eqref{eq:Dring} in main text).
The three panels summarize all studied systems.
Colorcode/symbols are as in Fig.~\ref{fig:PathLengthPDFs} in main text, with the legend here reporting the number of monomers per ring $N_{\rm ring}$.
Large fluctuations for $N_{\rm ring}=1280$ indicate that the corresponding trajectories are too short for determining $D_{\rm ring}$ with a precision analogous to the other cases.
}
\label{fig:g3Plot}
\end{figure}
%

\end{document}